\documentclass[12pt,preprint]{emulateapj}

\newcommand{\etal}{{et~al.\null}}
\newcommand{\eg}{{e.g.,}}

\newcommand{\ie}{{i.e.,}}
\newcommand{\kms}{km~s$^{-1}$}
\newcommand{\Rarrow}{\ \Longrightarrow\ \ }
\newcommand{\Oline}{[\ion{O}{3}]~$\lambda 5007$}
\newcommand{\simgt}{{\raise-.5ex\hbox{$\buildrel>\over\sim$}}\ }
\newcommand{\simlt}{{\raise-.5ex\hbox{$\buildrel<\over\sim$}}\ }

\newenvironment{packed_enum}{
\begin{enumerate}
  \setlength{\itemsep}{1pt}
  \setlength{\parskip}{0pt}
  \setlength{\parsep}{0pt}
}{\end{enumerate}}

\slugcomment{}

\shorttitle{Planetary Nebula Kinematics and Disk Mass}
\shortauthors{Herrmann \& Ciardullo}

\begin{document}

\title{Planetary Nebulae in Face-On Spiral Galaxies. \\ III. Planetary Nebula Kinematics and Disk Mass}

\author{Kimberly A. Herrmann\altaffilmark{1,2,3} and Robin Ciardullo\altaffilmark{1,2}}
\affil{Department of Astronomy \& Astrophysics, The Pennsylvania State University \\ 525 Davey Lab, University Park, PA 16802}
\email{herrmann@lowell.edu, rbc@astro.psu.edu}

\altaffiltext{1}{Visiting Astronomer, Cerro Tololo Inter-American Observatory (CTIO). CTIO is operated by the Association of Universities for Research in Astronomy, Inc.\ (AURA) under cooperative agreement with the National Science Foundation (NSF).}

\altaffiltext{2}{Visiting Astronomer, Kitt Peak National Observatory, National Optical Astronomy Observatories (NOAO), which is operated by AURA, Inc.\ under contract to the NSF.  The WIYN Observatory is a joint facility of the University of Wisconsin-Madison, Indiana University, Yale University, and NOAO.}

\altaffiltext{3}{Current address: Lowell Observatory, 1400 West Mars Hill Road, Flagstaff, AZ, 86001}

\begin{abstract}
Much of our understanding of dark matter halos comes from the assumption that the mass-to-light ratio ($\Upsilon$) of spiral disks is constant.  The best way to test this hypothesis is to measure the disk surface mass density directly via the kinematics of old disk stars.  To this end, we have used planetary nebulae (PNe) as test particles and have measured the vertical velocity dispersion ($\sigma_z$) throughout the disks of five nearby, low-inclination spiral galaxies:  IC~342, M74 (NGC~628), M83 (NGC~5236), M94 (NGC~4736), and M101 (NGC~5457).  By using H{\sc i} to map galactic rotation and the epicyclic approximation to extract $\sigma_z$ from the line-of-sight dispersion, we find that, with the lone exception of M101, our disks do have a constant $\Upsilon$ out to $\sim$3~optical scale lengths ($h_R$).  However, once outside this radius, $\sigma_z$ stops declining and becomes flat with radius.   Possible explanations for this behavior include an increase in the disk mass-to-light ratio, an increase in the importance of the thick disk, and heating of the thin disk by halo substructure.   We also find that the disks of early type spirals have higher values of $\Upsilon$ and are closer to maximal than the disks of later-type spirals, and that the unseen inner halos of these systems are better fit by pseudo-isothermal laws than by NFW models.

\end{abstract}


\section{INTRODUCTION}
Traditionally, spiral galaxies have three main components: a dynamically hot bulge, a dynamically cold disk (which can sometimes be divided into ``thin'' and ``thick'' components), and a mysterious dark matter halo.  However, measuring the mass of each component separately is a challenge, especially in a galaxy's extreme outer regions where the dark matter halo dominates.  Even in late-type spirals with minimal bulges, the mass profiles of the dark halos cannot be decoupled from the visible disk mass using rotation curves alone \citep{bsk04}.  Most analyses have therefore relied on the ``maximal disk'' hypothesis, wherein the disk mass-to-light ratio ($\Upsilon$) is assumed to be constant with radius, and as large as possible to account for the rotation in the inner parts of the galaxy \citep[\eg][]{kent86, pw00, sofue03}.

Is the disk $\Upsilon$ really constant and maximal? Perhaps the best way to break the disk-halo degeneracy is to measure the surface mass of a disk directly from the $z$~motions of stars.  In systems where the primary stellar motion is circular, the phase-space distribution of stars in the direction perpendicular to the plane is fixed by the disk's gravitational potential \citep{bt87}.  For example, in the Milky Way, star counts and velocity measurements toward the Galactic pole initially suggested that the stellar disk is roughly isothermal \citep[\eg][]{oort}.  If this were indeed the case, then the integrated face-on stellar velocity dispersion, $\sigma_z$, would be related to the disk surface mass, $\Sigma$, by
\begin{equation}
\sigma_z^2(R) = K G \, \Sigma(R) h_z,
\label{isothermal}
\end{equation}
where $K = 2 \pi$ is a constant, $G$ is the gravitational constant, and  $h_z$ is the scale height of the stars \citep{vdK88}.   Since surveys of edge-on galaxies \citep[\eg][]{vdKs82, bm02} imply that $h_z$ is roughly constant with radius, measurements of $\sigma_z$ can produce estimates of disk mass that are independent of a galaxy's rotation curve.

Absorption line studies have shown that, in the inner regions of spirals, $\sigma_z$ generally follows the exponential law expected from a constant $\Upsilon$ disk \citep{b93, g+97, g+00}.  However, these surveys were limited by surface brightness, and do not extend more than $\sim$2~scale lengths from the nucleus.  To probe the outer regions of disks, where the dark matter halo is more important, some other technique is required.  Planetary nebulae (PNe) are ideal tools for this purpose: PNe are extremely bright in [\ion{O}{3}], abundant out to $\gtrsim$5~disk scale lengths, measurable to $\sim$2~\kms\ with fiber-fed spectrographs, and present in all stellar populations with ages between $10^8$ and $10^{10}$~yr. Moreover, these old disk stars are easy to distinguish from other emission-line sources via their distinctive [\ion{O}{3}]-H$\alpha$ ratio \citep{p12}. 

\begin{deluxetable*}{cccccccccc}
\tabletypesize{\scriptsize}
\tablecaption{Program Galaxies\label{tabBasic}}
\tablewidth{0pt}
\tablehead{&\colhead{Hubble}&&\colhead{Distance}&\colhead{$h_R$}&\colhead{Disk}&&\colhead{$v_{max}$}&\colhead{Number of}&\colhead{Survey} \\
\colhead{Galaxy} &\colhead{Type\tablenotemark{a}} &\colhead{$i$\tablenotemark{b}} &\colhead{(Mpc)\tablenotemark{c}} &\colhead{(kpc)\tablenotemark{d}} &\colhead{$\mu_0$\tablenotemark{e}} &\colhead{$E(B-V)$\tablenotemark{f}} &\colhead{(\kms)\tablenotemark{g}} &\colhead{PN Velocities} &\colhead{Region}
}
\startdata
IC~342 &Scd  &$31^\circ$  &$3.5 \pm 0.3$ &4.24 &19.60 &0.558 &200 &99 &$4\farcm 8$  \\
M74    &Sc   &$6.5^\circ$ &$8.6 \pm 0.3$ &3.17 &20.20 &0.070 &170 &102 &$4\farcm 8$ \\
M83    &SBc  &$24^\circ$  &$4.8 \pm 0.1$ &2.45 &19.07 &0.066 &255 &162 &$18\arcmin$ \\
M94    &Sab  &$41^\circ$  &$4.4^{+0.1}_{-0.2}$ &1.22  &18.88 &0.018 &200 &127 &$5\farcm 8$ \\
M101   &Scd  &$17^\circ$   &$7.3 \pm 0.5$ &4.99 &20.29 &0.009 &250 &60 &$8\arcmin$
\enddata
\tablenotetext{a}{From RC3}
\tablenotetext{b}{IC 342: \citet{cth00}; M74: \citet{fb+07}; M83: \citet{p+01}; M94: \citet{dB+08}; M101: \citet{ZEH90}} 
\tablenotetext{c}{From Paper~I, except for M101, which is from \citet{M101PNe}; note that this distance has been updated to be consistent with $M^* = -4.47$ \citep{p12}}
\tablenotetext{d}{IC~342: \citet[][$I$-band]{wb03}; M74: \citet[][$R$-band]{m04}; M83: \citet[][$R$-band]{k+00}; M94: \citet[][$3.6\mu$m]{dB+08} and \citet[][$R$-band]{g+09}; M101: \citet[][$R$-band]{khc06}}
\tablenotetext{e}{In mag~arcsec$^{-2}$, corrected for galactic inclination. IC~342: \citet[][$I$-band]{wb03}; M74: \citet[][$R$-band]{m04}; M83: \citet[][$R$-band]{k+00}; M94: \citet[][$R$-band]{g+09}; M101: \citet[][$R$-band extrapolation]{k+00}}
\tablenotetext{f}{From \citet{sfd98}}
\tablenotetext{g}{IC 342: \citet{cth00}; M74: \citet{fb+07}; M83: \citet{p+01}; M94: \citet{dB+08}; M101: \citet{ZEH90}}
\end{deluxetable*}

Planetary nebulae have only recently been used to probe the kinematics of spiral galaxies.  The largest study of this kind has been for the Local Group galaxy M31, where the analysis of over 2000 PN velocities has demonstrated a flattening of the disk's line-of-sight velocity dispersion at large galactocentric radii \citep{M31PNS}.  Although the high-inclination of the galaxy precluded any dynamical measurement of disk mass, these data did present evidence for disk flaring at $R \gtrsim 11$~kpc.  The analysis of 140~PN velocities in the moderately inclined ($i \sim 56^\circ$) Local Group spiral M33 yielded a similarly surprising result:  by assuming a constant scale-height for the galaxy's disk, \citet{M33PNe} concluded that the disk mass-to-light ratio actually increases with radius, going from $\Upsilon_V \sim 0.3$ in the inner regions to $\Upsilon_V \sim 2.0$ at $\sim 9$~kpc.  Finally, \citet{M94PNS} measured the velocities of 67 PNe in the low-inclination ($i \sim 35^\circ$) spiral M94, and found a line-of-sight velocity dispersion that declines exponentially over $\sim$4~disk scale lengths.

Papers~I and II of this series \citep{thesis1, thesis2} presented narrow-band imaging and follow-up spectroscopy of PNe in a sample of large ($r > 7\arcmin$), nearby ($D < 10$~Mpc), low-inclination ($i \leq 41^\circ$) spiral galaxies.  Here we use these data to understand the disk kinematics of five galaxies:  IC~342, M74, M83, M94, and M101.  We introduce our sample in \S2, and in \S3, we discuss the problem of the disk scale height, the one parameter of our analysis that cannot be measured.   We explore its dependence on galactic radius, its relationship to other disk properties, such as scale length, and alternatives to the isothermal disk assumption.  We then begin our analysis in \S4 by compensating for the fact that none of our nearby galaxies is precisely face-on.  We use H{\sc i} velocity maps to deproject the galactic disks, define the circular velocity as a function of radius, and estimate asymmetric drift.  Next, in \S5 we identify those PNe which likely belong to the galaxies' spheroidal components, and eliminate them from the analysis.  We find that very few objects fall into this category: only three in IC~342, four in M74, six in M83, three in M94, and one in M101.  Then, in \S6, we examine the PNe's line-of-sight dispersions without these halo objects.   We show that in the inner regions of our galaxies, the line-of-sight velocity dispersion, $\sigma_{los}$, generally declines as expected for a constant $h_z$, constant $\Upsilon$ disk.  However, once past $\sim$4~optical scale lengths, $\sigma_{los}$ flattens out at a value much larger than expected.   We consider a few possible explanations for this phenomenon, including the presence of emission-line contaminants in our sample, the effects of internal extinction, and the behavior of large-scale galactic warps.  In \S7, we model the disks' velocity ellipsoids, and extract $\sigma_z$ from the line-of-sight velocity dispersions.  As expected, we find that $\sigma_R$ and $\sigma_{\phi}$ contribute little to the measured dispersion of our nearly face-on disks.  In \S8, we describe the results for each of the five galaxies in our study.  We convert our values of $\sigma_z$ into disk surface masses and calculate disk mass-to-light ratios.  Then, in \S9, we analyze the results from the inner and outer regions of the galaxies.  We compute disk and halo rotation curves, compare our disk mass surface densities to those expected from maximal disks, show that pseudo-isothermal cores fit the residual rotation curves better than NFW models, and discuss the physical mechanisms that may produce a constant $\sigma_z$, including the interaction of thin and thick disks with halo substructure.   We conclude in \S10 by summarizing our results and suggesting future observations that will assist in their interpretation.

\section{THE SAMPLE}
Table~1 lists the five galaxies observed in this survey, along with their physical properties.   IC~342, M74, and M101 are large, late-type spirals with little or no spheroidal components, and rotation velocities greater than 170~\kms.  M94 is an earlier system (type Sab) with a small, but bright bulge; in this galaxy, our PN identifications were largely confined to regions outside of $\sim$1.5~kpc, where the disk of the galaxy dominates.  Finally, M83 is a strongly barred SBc system with isophotal twists and a warped outer disk.

Our galaxies were initially surveyed for planetary nebulae with on-band/off-band [\ion{O}{3}] and H$\alpha$ imaging with the WIYN telescope in the north and the CTIO Blanco telescope in the south (Paper~I).   Follow-up spectroscopy was then accomplished via the Hydra multi-object spectrographs of these same two telescopes (Paper~II).  These observations produced between 60 and 162 PN radial velocities per galaxy, with a median measurement error of $\sigma_v \sim 6$~\kms, and, in all cases, $\sigma_v < 15$~\kms.   The positions, magnitudes, radial velocities, and velocity uncertainties of all the PN candidates are given in Paper~II.

\begin{deluxetable*}{cccccccc}
\tabletypesize{\scriptsize}
\tablecaption{Scale heights (in pc)\label{tabhz}}
\tablewidth{0pt}
\tablehead{\colhead{Galaxy} &\colhead{dG\tablenotemark{a}} &\colhead{K02\tablenotemark{b}} &\colhead{BM02\tablenotemark{c}} &\colhead{Initial Range} &\colhead{Stability\tablenotemark{d}} &\colhead{Rotation Curve\tablenotemark{e}} &\colhead{Final Range} 
}
\startdata
IC 342 &505  &424  &565  &$400-600$ &$>300$  &$>100$  &$400-600$ \\
M74    &369  &327  &423  &$300-500$ &$>300$  &$>170$  &$300-500$ \\
M83    &285  &377  &402  &$200-500$ &$>300$  &$>200$  &$300-500$ \\
M94    &277  &186  &218  &$100-300$ &$>300$  &$>200$  &$300-400$ \\
M101   &623  &523  &697  &$500-700$ &$>300$  &$>60$  &$500-700$ 
\enddata
\tablenotetext{a}{Estimated using $h_R/h_z$ values from \citet{dG98}}
\tablenotetext{b}{Estimated using $h_R/h_z$ values from \citet{kvd02}}
\tablenotetext{c}{Estimated using $h_R/h_z$ values from \citet{bm02}}
\tablenotetext{d}{Stability limit on $h_z$}
\tablenotetext{e}{Rotation curve limit on $h_z$}
\end{deluxetable*}

\section{ESTIMATING SCALE HEIGHT}
Before we can translate our PN velocities into disk mass, we need to adopt some form for the phase-space distribution of stars perpendicular to the disk.   This is not an entirely settled question, even in the Milky Way.  As described by \citet{vdK88}, the vertical structure of stars in a galactic disk can be parameterized by a family of models 
\begin{equation}
\rho(z) \propto {\rm sech}^{2/n} \left( {n z \over 2 h_z} \right)
\label{family}
\end{equation}
which have isothermal ($n = 1$) and exponential ($n = \infty$) distributions
as their limiting cases.  Depending on the value of $n$, Poisson's
equation yields different constants for the relationship between the disk
mass surface density and the integrated face-on velocity dispersion.  
In other words, in equation~(\ref{isothermal}),
\begin{eqnarray}
n = 1  \Rarrow& \rho_{\rm iso}(z)  \propto {\rm sech}^2 (z /2 h_z) &\Rarrow K = 2 \pi \\
n = 2   \Rarrow& \rho_{\rm im}(z)   \propto {\rm sech} (z / h_z)    &\Rarrow K = 1.7051 \pi \nonumber \\
n = \infty  \Rarrow& \rho_{\rm exp}(z)  \propto \exp(-z / h_z)          &\Rarrow K = 1.5 \pi. \nonumber 
\label{cases}
\end{eqnarray}
\Citet{vdK88} has argued that the intermediate ($\rho_{\rm im}$) case best fits star counts in the Milky Way, especially at low $z$, where the superposition of many stellar populations, each with their own value of $\sigma_z$, blend together.  To take this blending into account, \citet{siebert} combined several isothermal components to find a result similar to the exponential case.  However, when initial mass functions and star formation rates were considered, \citet{haywood} found that a single exponential was insufficient to model the Galaxy's vertical structure.  More recently, \citet{sd00} have shown that in a sample of 61 edge-on spirals, 60\% have vertical distributions that are best fit by the sech$(z)$ law.   However, unlike the Milky Way star counts, these measurements were the result of a fit to surface brightness, and therefore represent a solution that is weighted by luminosity (actually, by the inverse of the stellar mass-to-light ratio), rather than the actual mass density of stars.

While PN observations do not strictly sample mass, neither do they produce results that are luminosity-weighted.  Generally speaking, PNe are excellent tracers of light, but only in populations where the mass-to-light ratio changes slowly with age, \ie\ in systems older than $\sim 0.1$~Gyr \citep{maraston}.  Younger populations, which may dominate the luminosity of a galaxy, produce no planetary nebulae whatsoever.  Thus, the distribution of PNe should be closer to one that is space-weighted, rather than one that is luminosity weighted.

Given these issues we chose to model our disks using an intermediate ($n=2$) vertical distribution.  This assumption increases our derived disk masses by 17\% over the isothermal case, while decreasing it by $\sim 14\%$ compared to exponential distributions.

With the vertical distribution set, we next needed to establish the values of $h_z$ appropriate for our sample of galaxies.  Since all of our targets are low-inclination systems, their vertical scale heights are not directly observable.  Consequently, in order to constrain $h_z$, we had to use correlations with other measurable quantities, such as disk scale length ($h_R$) and disk surface brightness.  Several studies have investigated these trends, generally finding $h_R/h_z \sim 10$ for large, non-interacting late-type spirals, such as those observed in our survey \citep[\eg][]{dGvdK96, kvd02, yd06}.  In addition, \citet{dG98} and \citet{kvd02} have explored the behavior of $h_R/h_z$ versus Hubble type.  Both studies find that, though later-type spirals generally have larger $h_z$ values than earlier type objects, they also have larger values of $h_R/h_z$.  Thus, while Scd galaxies have $h_R/h_z \sim 10$, the Sab systems have ratios as low as $\sim$5.  Finally, \citet{bm02} have examined the dependence of $h_R/h_z$ as a function of $K$-band central surface brightness, $K_0$, finding
\begin{equation}
h_R/h_z = \left( 0.65 - 0.0295 K_0 \right)^{-1}.
\label{2MASSEq}
\end{equation}
This relation has a substantial amount of scatter, but if we use the radial surface brightness profiles of the 2 Micron All Sky Survey (2MASS) Large Galaxy Atlas website\footnotemark[4] to estimate $K_0$, then the implied values of $h_R/h_z$ for the galaxies in our sample range from $\sim$5.6 to $\sim$7.5.

\footnotetext[4]{http://www.ipac.caltech.edu/2mass/gallery/largegal/atlas.html}

Table~\ref{tabhz} summarizes our initial estimates for $h_z$, based on the galactic scale lengths given in Table~1.  In \S7, we will further investigate the plausibility of these numbers using stability arguments applied to the observed PN kinematics.  We will also consider whether the use of a single value of $h_z$ is appropriate, or whether ``flaring'' must be considered in the extreme outer disks of galaxies.

\begin{figure*}
\epsscale{0.85}
\plotone{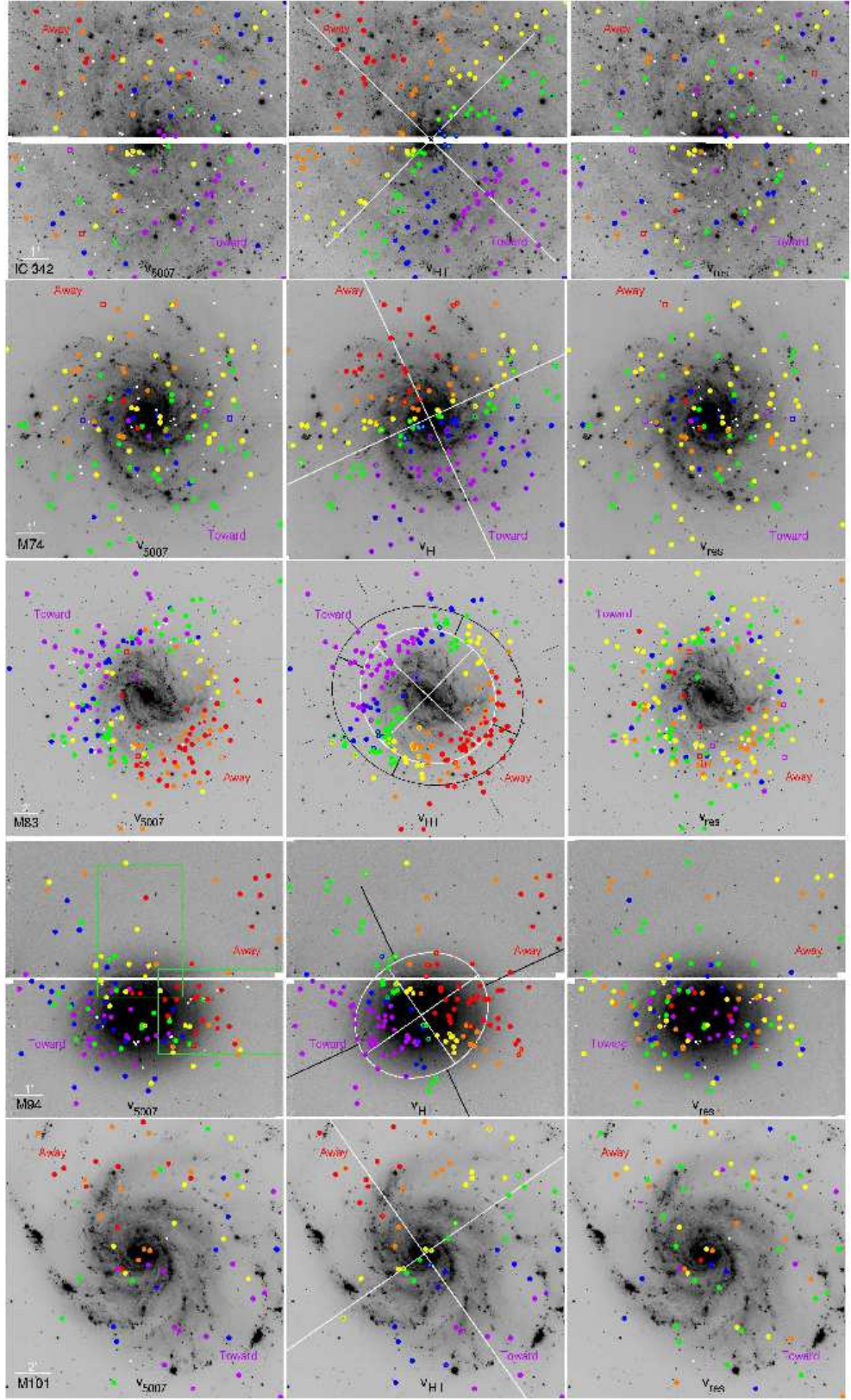}
\caption{\scriptsize [O~III] $\lambda 5007$ images of our program galaxies, with the positions of the PNe superposed.  {\it Left:} observed PN radial velocities. {\it Middle:} VLA H{\sc i} velocities (outer annuli) and our model of galactic rotation (inner dots). {\it Right:} velocity residuals formed by subtracting the VLA H{\sc i} velocities (corrected for asymmetric drift) from our PN velocities.  Color coding for all but the M74 H{\sc i} figure is in 30~\kms\ increments from $> 60$~\kms\ systemic (red) to $< - 60$~\kms\ systemic (violet).  H{\sc i} velocities for the face-on M74 system are shown in 10~\kms\ increments, from $20 < v < 30$~\kms\ to $-30 < v < -20$~\kms.  Solid points are well-determined velocities ($\sigma_v < 15$~\kms), crosses are uncertain velocities ($\sigma_v > 15$~\kms), squares represent probable halo objects, and small white dots denote photometrically identified PNe with no velocity measurement.   The green boxes in the left panel of M94 show the fields studied by \citet{M94PNS}.  The axes in the middle panels illustrate our adopted values for the galactic position angles.  North is up and east is to the left.  \label{onbands} }
\end{figure*}

\section{VELOCITY FIELDS FROM PLANETARY NEBULAE}
\begin{figure}[b]
\epsscale{1.15}
\plotone{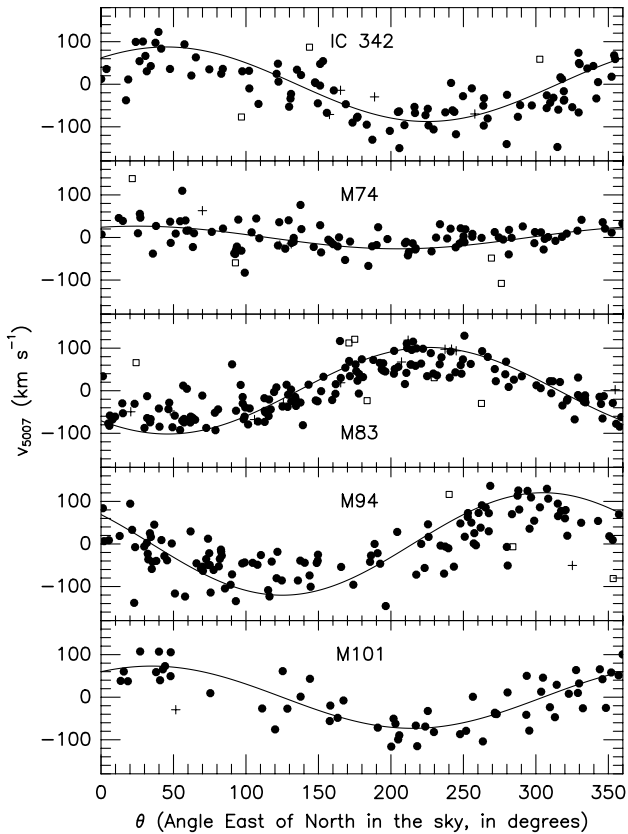}
\caption{\scriptsize PN radial velocities as a function of position angle on the sky, where $\theta$ is the angle East of North.   The sinusoidal pattern illustrates galactic rotation, and the location of maximum velocity, $v_{max} \sin i$, defines the position angle of the major axis.  Solid points are well-determined velocities ($\sigma_v < 15$~\kms), crosses are uncertain velocities ($\sigma_v > 15$~\kms), and squares represent probable halo objects. \label{v5007theta} }
\end{figure}

The line-of-sight velocity ($v_{los}$) of a planetary nebula rotating in the disk of a galaxy is given by
\begin{equation}
v_{los} = v_{\phi} \cos \phi \sin i + v_R \sin \phi \sin i + v_z \cos i + v_{sys},
\label{vlos_eq}
\end{equation}
where $v_{sys}$ is the galactic systemic velocity, $i$ the disk inclination, and $v_R$, $v_{\phi}$, and $v_z$ are the PN's component velocities in the radial ($R$), azimuthal ($\phi$), and vertical ($z$) directions.  Note that even though our galaxies are nearly face-on, the effect of rotation is still considerable.  This is easily seen in the left panels of Figure~\ref{onbands}, which display our measured PN velocities superimposed on an \Oline\ image of each galaxy.  Clearly $v_{\phi}$ is much larger than $v_z$ even when it is diminished by $\sin i$.  (The one exception is M74 which has an inclination of only $\sim$6$^\circ$.)  Figure~\ref{v5007theta} shows the rotation in another way by plotting our measured velocities as a function of azimuthal angle, $\theta$ (north through east in the plane of the sky).  The location of the maximum of each sinusoid indicates the position angle of the principal axis, while the amplitude of the curve places a limit on $v_{max} \sin i$.  This rotation must be removed before $\sigma_z$ can be isolated.  Fortunately, The H{\sc i} Nearby Galaxy Survey \citep[THINGS;][]{THINGS} collaboration has recently used the Very Large Array (VLA) to create very high resolution velocity maps of four of our five spirals.  Moreover, \citet{IC_HI} have made a similar H{\sc i} map for IC~342, though with lower resolution.

\subsection{Removing Galactic Rotation}
To remove galactic rotation from our PN velocities, we first needed to define each disk's three dimensional geometry, and deproject the coordinates from the plane of the sky to the plane of the galaxy.  We started with literature values for each galaxy's position angle, inclination, and rotation curve, and then iteratively refined the parameters to fit the VLA H{\sc i} velocity maps.  In some cases, this meant changing the quoted position angle of the major axis by a few degrees; at other times, the inclination or rotation amplitude required modification.  When the latter condition occurred, we preferred to tweak the inclination by less than five degrees rather than adjust the rotation velocity by 20 to 60~\kms.  Table~\ref{tabParams} presents our final geometric parameters for our disks.

\begin{deluxetable}{cccc}
\tabletypesize{\scriptsize}
\tablecaption{Geometric Parameters for Disk Models\label{tabParams}}
\tablewidth{0pt}
\tablehead{\colhead{Galaxy} & \colhead{PA} & \colhead{$i$} & \colhead{Radial Range}\\
							& (in degrees) & (in degrees)  & (in arcmin)}
\startdata
IC~342	& 45		& 31		&Full range \\
M74		& 25		&  9		&Full range \\
M83		&226		& 24		&$R < 5.8$ \\
		&246		& 27		&$5.8 - 8.0$ \\
		&$236-226$	&$29-30$	&$8.0 - 9.0$\tablenotemark{a} \\
		&$226-201$	&$30-35$	&$9.0 - 14.0$\tablenotemark{a} \\
		&$201-172$	&$35-46$	&$14.0 - 25.0$\tablenotemark{a} \\
M94		&305		& 37		&$R < 2.45$ \\
		&295		& 37		&$R > 2.45$ \\
M101	& 35		& 17		&Full range
\enddata
\tablenotetext{a}{PA and $i$ are linearly dependent on $R$ in these regimes. (Based on model in \citet{M83HI}.)}
\end{deluxetable}

The middle panels of Figure~\ref{onbands} compare our final models for galactic rotation with the measured H{\sc i} velocity at the location of the objects tabulated in Paper~II.  For IC~342, M74, and M101, a simple thin disk with a constant position angle and inclination works well over the entire region of our study.  For the more extended survey of M94, two different position angles were needed, one for the inner disk (PA = $305^\circ$), and one for the system's outer ring (PA = $295^\circ$).  Finally, to model the known isophotal twists of M83, we started with the warped disk model of \citet{M83HI}, which extends well beyond our survey area, and then adjusted the parameters to fit the higher resolution THINGS data.  The axes in the middle panels of Figure~\ref{onbands} illustrate our final values for the galactic position angles.  

We do note that just because our H{\sc i} residuals exhibit no systematic trend with azimuth or radius, that does not mean that a map of the stellar velocity would be just as well-behaved.   However, it is likely that the two are linked.  \citet[][and references therein]{m+97} have found that disks which contain warps in H{\sc i} often have stellar warps as well.  Moreover, in at least one case (NGC~5907), both warps have similar orientations \citep{m+94}.  Thus, our procedure of modeling the disk orientation using H{\sc i} data should be valid.

\subsection{Estimating Asymmetric Drift}
\begin{figure}
\epsscale{1.1}
\plotone{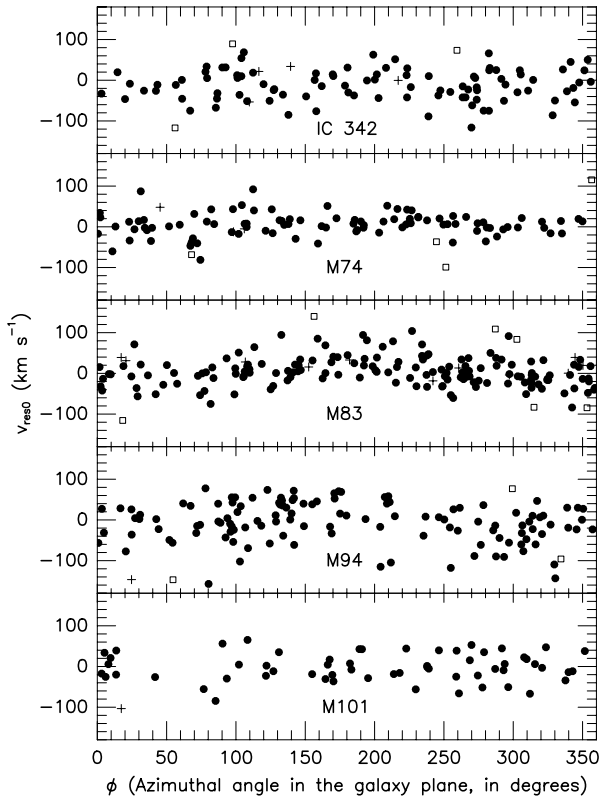}
\caption{\scriptsize PN velocities after the removal of disk rotation, as defined by the motion of the H{\sc i} gas.  Note that for M83 and M94, the pattern of velocity residuals has a slightly convex appearance.  This is the signature of velocity oversubtraction, due to the action of asymmetric drift.  See Figure~\ref{v5007theta} for an explanation of the point types. \label{vres0phi} }
\end{figure}

Once the geometry of the disk was set, we subtracted off the H{\sc i} rotation from the observed PN velocities.  This generated the velocity residuals shown in Figure~\ref{vres0phi}.  As the figure illustrates, the H{\sc i} rotation curves do not remove all trace of systematic behavior from the velocities.  In particular, the PN velocities of M83 and M94 now show a low-amplitude convex bowing with respect to $180^\circ$ azimuth.  This is the signature of asymmetric drift, \ie\ the lag between the circular velocity, as measured by the gas, and the average azimuthal velocity of the stars.   By using the H{\sc i} data to define rotation, we oversubtracted the stars' mean azimuthal motion.

To estimate the amplitude of this effect, we can use the velocities of the PNe themselves.  Because galactic disks are cold systems, the $v_{\phi} \cos \phi \sin i$ term of equation~(\ref{vlos_eq}) clearly dominates.  Thus, to a first approximation,
\begin{equation}
v_{\phi} \approx \frac{v_{los} - v_{sys}}{\cos \phi \sin i}.
\end{equation}
Obviously, at $\phi = 90^\circ$ and $\phi = 270^\circ$ the solution to this equation is undefined, but if we exclude PNe within $10^\circ$ of the minor axis and bin by radius, we can derive mean values of $v_{\phi}$ over the entire disk of the galaxy.  These values can then be compared to the observed motion of the gas.  This analysis confirms the results of Figure~\ref{vres0phi}: that asymmetric drift is negligible in IC~342 and M101, $\sim$15~\kms\ in M74, between 20 and 40 \kms\ in M83, and between 40 and 50 \kms\ in M94.

Are these velocities reasonable?  It is important to note that asymmetric drift is extremely difficult to measure, so very few estimates exist in the literature.  In the solar neighborhood, where the rotational velocity is $v_{max} \sim 220$~\kms, the asymmetric drift is $v_{asd} \sim 20$~\kms\ \citep{db98, od03}.  This is roughly twice the value found by \citet{M33PNe} from radial velocity measurements of PNe in the small, late-type spiral M33 ($v_{max} \sim 100$~\kms).  For the large Sb spiral M31 ($v_{max} \sim 260$~\kms), PN measurements yield values of $v_{asd}$ between 40 to 50~\kms\ \citep{M31PNS}, and in the barred lenticular NGC~1023 ($v_{max} \sim 220$~\kms), $v_{asd} \sim 40$~\kms\ \citep{n+08}.  These estimates suggest that asymmetric drift is usually 10-20\% of the rotation velocity in late-type galaxies and slightly larger in earlier systems.  Both correlations make sense:  in the Milky Way, asymmetric drift is correlated with stellar population, since older stars have had more time to scatter \citep{db98}, and the amount (and amplitude) of this scatter depends on the speed of rotation.

\begin{deluxetable}{cccccc}
\tabletypesize{\scriptsize}
\tablecaption{Asymmetric Drift Estimates (\kms)\label{tabASD}}
\tablewidth{0pt}
\tablehead{\colhead{Galaxy} &\colhead{Estimate} &\colhead{$v_{max}$} &\colhead{$0.1 v_{max}$} &\colhead{$0.2 v_{max}$} &\colhead{Best Value} }
\startdata
IC 342       &negligible      &200     &20     &40     &20 \\
M74          &15              &170     &17     &34     &34 \\
M83          &$20-40$         &255     &26     &51     &51 \\
M94          &$40-50$         &200     &20     &40     &40 \\
M101         &negligible      &250     &25     &50     &25
\enddata
\end{deluxetable}

Table~\ref{tabASD} compares our observed asymmetric drift velocities to values which might be inferred from a simple scaling of the rotation curve. To facilitate this comparison, we have used a single drift value for each galaxy; this is consistent with the PN data of M31, M33, and NGC~1023, which suggest that the asymmetric drift within a galaxy is roughly constant with radius.  We also list the ``best fit value'' for each object, \ie\ the one that minimizes our residual line-of-sight PN velocity dispersions while still being in the range between 0.1 and $0.2 \, v_{max}$ (see next section).   As Figure~\ref{vres0phi} and Table~\ref{tabASD} indicate, our galaxies exhibit a range of drift velocities.  M83, which is a strongly barred galaxy and the fastest rotator in our sample, has the largest value of $v_{asd}$.  In M94, which is an early-type system (Hubble type Sab), $v_{asd}$ is also rather large, $\sim$40~\kms.  On the other hand, IC~342 and M101 both have rather low values for the asymmetric drift.  This suggests that their disks are relatively young, an idea supported by the fact that these two galaxies are the latest systems in our sample (Hubble types Scd).  Finally, the asymmetric drift for M74 is slightly higher than expected; however, due to the galaxy's extremely low inclination ($\sim 6^\circ$), this number is very uncertain.  We will revisit our estimates of $v_{asd}$ in \S 9, when we have measurements for the disks' radial and perpendicular velocity dispersions.

The right panels of Figure~\ref{onbands} display our PN line-of-sight velocities with the H{\sc i} gas rotation and our best-fit asymmetric drift values removed.  The data show no trend with $\phi$, indicating that the adopted drift values are reasonable.  We note that the results that follow depend very little on the exact values of $v_{asd}$: because our galaxies all have low-inclinations, 10 to 20~\kms\  changes in the asymmetric drift have almost no effect on the final values for the $z$ component of the velocity dispersion.

\begin{figure}[t]
\epsscale{1.16}
\plotone{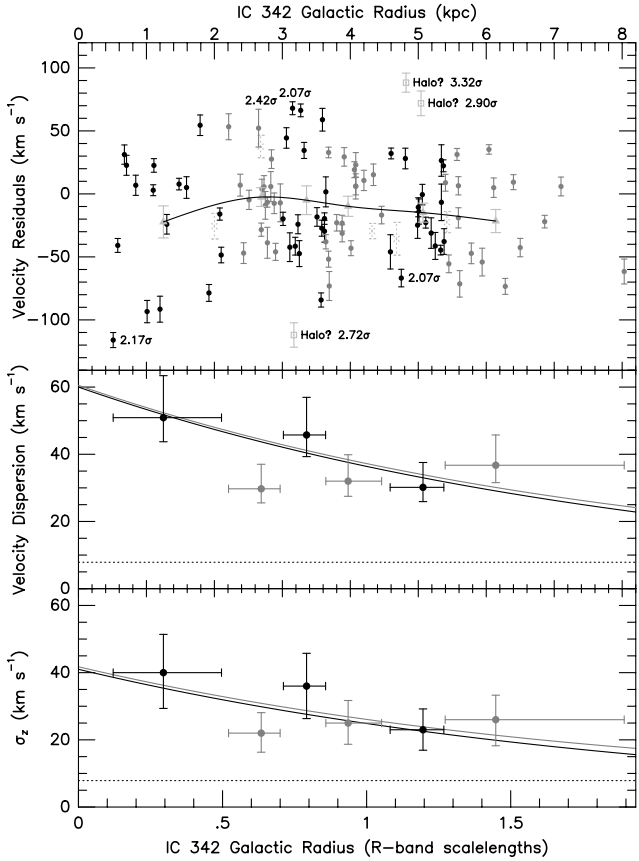}
\caption{\scriptsize IC~342 PN radial velocities ({\it top}), line-of-sight velocity dispersion ({\it middle}), and velocity dispersion in the $z$ direction ({\it bottom}), all as a function of galactic radius.  PNe rejected as non-disk objects are identified in the top panel, along with the amount of their departure from the mean disk; these objects are not included in the lower plots.  PNe with velocity uncertainties greater than 15~\kms\ are not shown; some H{\sc ii} region contaminants are plotted as light gray crosses.  The light gray triangles and black curve in the top panel indicate the average residual velocity in each bin.  The alternating black and dark gray points illustrate the radial bins. In the two lower panels, the dotted line shows our median velocity error, the solid black curve denotes an exponential with twice the galaxy's $I$-band scale length, and the gray curve adds these two relationships in quadrature.  This last curve is the one expected for a constant $h_z$, constant $\Upsilon$ disk.  The radial width of each 16 object bin is illustrated via the errors bars in $x$; 1$\sigma$ velocity uncertainties are indicated by the errors in $y$.  For IC~342, the velocity dispersion agrees with the simple model of a constant $h_z$, constant $\Upsilon$ disk. \label{ICrsds3} }
\end{figure}

\section{IDENTIFICATION OF HALO OBJECTS}
After the removal of rotation, our next step was to exclude those objects not part of the general disk population.  Since all five of our program galaxies are disk-dominated systems, the vast majority of the PNe should be rotating close to the galactic plane.  However, a few halo objects may be present, and these objects can greatly distort our measurement of the disk velocity dispersion.   For example, a halo PN will typically orbit at a velocity close to $v_{max}$, but with some random orientation in the galaxy.  For a nominal rotation speed of $\sim$200~\kms, this implies a velocity of $\sim$100~\kms\ with respect to the local standard of rest.  This is much larger than the expected value for velocity dispersion of disk stars.

\begin{figure}[b]
\epsscale{1.16}
\plotone{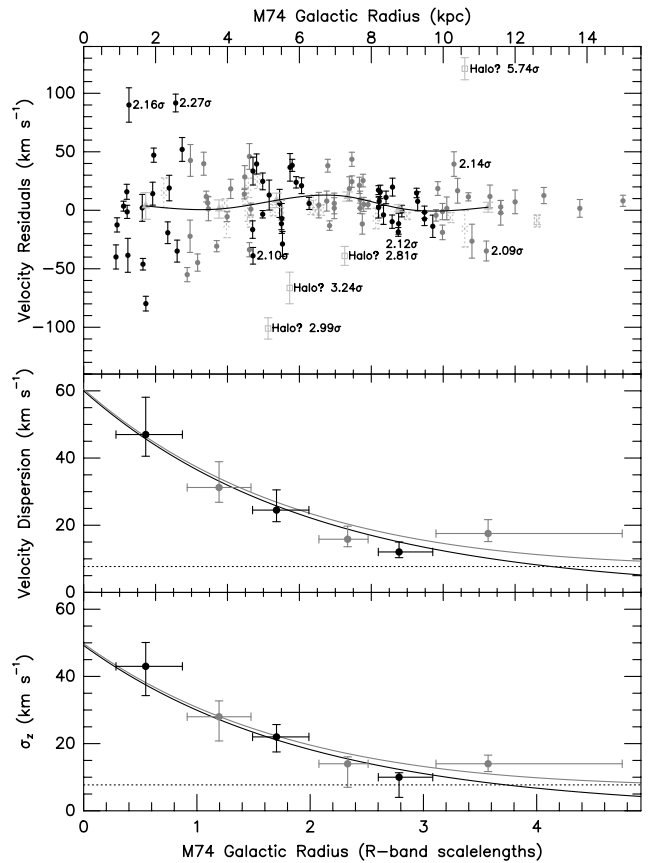}
\caption{\scriptsize M74 PN radial velocities ({\it top}), line-of-sight velocity dispersion ({\it middle}), and velocity dispersion in the $z$ direction ({\it bottom}), all as a function of galactic radius.  See the caption of Figure~\ref{ICrsds3} for more information.  Each radial bin contains 16-17 objects.  As with IC~342, the dispersion curve agrees with that expected for a constant $h_z$, constant $\Upsilon$ disk.  Our value of $53 \pm 5$~\kms\ for the central $z$ velocity dispersion is consistent with the value of $\sim$60~\kms\ derived from integral-field spectroscopy of the system's central kiloparsec \citep{g+06}. \label{M74rsds3}}
\end{figure}

Still, identifying halo PNe can be difficult.  Depending on the orientation of its orbit, a halo PN can have a line-of-sight velocity that mimics that of the disk.  More importantly, if the disks of our program galaxies are anything like that of the Milky Way, we can expect a line-of-sight velocity dispersion of $\sim$40~\kms.  Thus a typical halo planetary may have a velocity that is only $\sim$2.5~$\sigma$ away from the mean of the disk population.

Nevertheless, we can identify obvious halo candidates via a sigma-clipping algorithm.  For each PN, we identified its 15 closest neighbors (in radial distance) and, using those objects, we estimated the local disk velocity dispersion.   Any PN more than 2.5 standard deviations away from the mean (where $\sigma$ is defined from the disk dispersion and PN velocity uncertainty added in quadrature), was flagged as a possible halo contaminant and eliminated from the analysis.  Formally, this means that out of our entire sample of $\sim$550~PNe, we may be improperly excluding as many as $\sim$11~disk objects.  However, given the deleterious effect that halo PNe may have on our results, we feel that $2.5 \, \sigma$ is a reasonable threshold for PN exclusion.

\begin{figure}[t]
\epsscale{1.2}
\plotone{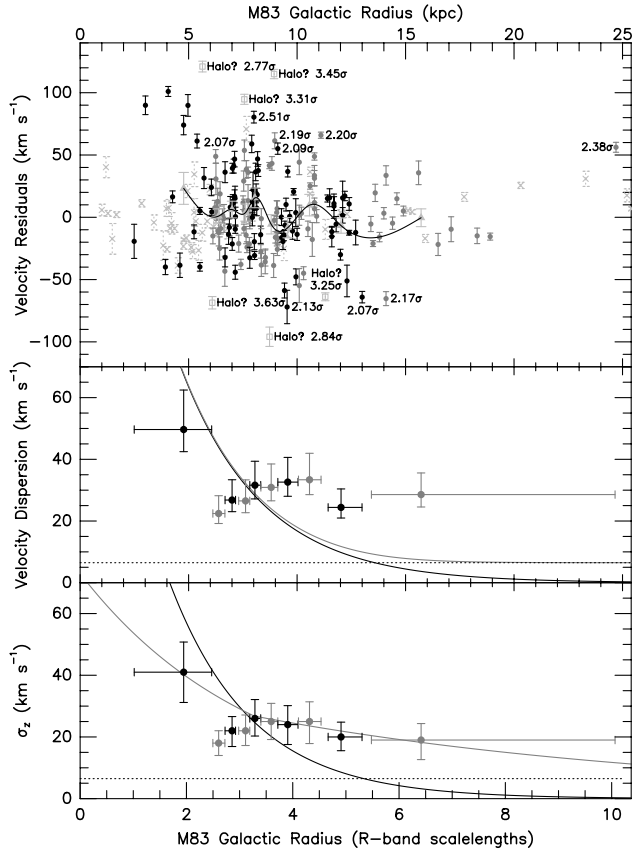}
\caption{\scriptsize M83 PN radial velocities ({\it top}), line-of-sight velocity dispersion ({\it middle}), and velocity dispersion in the $z$ direction ({\it bottom}), all as a function of galactic radius.  See the caption of Figure~\ref{ICrsds3} for more information.  Each radial bin contains 15-16 objects.   The gray line in the lower panel illustrates our best-fit two-component model for the galaxy.  M83 has a strong bar, and the small values of $\sigma_{los}$ and $\sigma_z$ in the second and third bins may be due to the bar's outer Lindblad resonance, which is located at $R \sim 7$~kpc.  If these two points are excluded from the analysis, then the dispersion profile looks similar to that of M94, with an exponentially declining inner region, and an asymptotically flat outer disk. \label{M83rsds3} }
\end{figure}

The top panels of Figures~\ref{ICrsds3}-\ref{M101rsds3} show the velocity residuals of all our PNe and identify those objects with discrepant velocities.  Even with our harsh $2.5 \, \sigma$ condition, very few PNe are excluded from the analysis.  In M74, PN~127 is an obvious ($\sim$6$~\sigma$) outlier and PNe~66, 96, and 146 also appear to be contaminating halo objects.  In IC~342, PNe~42, 98, and 111 are likely halo interlopers, while in M83 PNe~8, 43, 48, 50, 102 and 175 have the kinematics of the spheroidal component.  Despite the fact that M94 is an Sab galaxy, only three of its PNe (26, 63, and 102) are questionable: this is likely due to the fact that the galaxy's bulge is small, and that our PN identifications were largely confined to the galaxy's outer regions where the disk dominates.  Finally, none of the PNe in M101's kinematic sample have discrepant velocities: although PN~51 is a possible halo object, we had already excluded it from the analysis due to its large measurement error.

\section{RESIDUAL VELOCITIES AND THE LINE-OF-SIGHT DISPERSION}

With rotation removed and the halo PNe eliminated from the analysis, we could examine the kinematics of the galactic disks.  We started by assigning an equal number of PNe per radial bin (16 per bin in IC~342, M74, and M83, 18 per bin in M94, and 15 per bin in M101), and computing the line-of-sight velocity dispersion within each bin.  According to equation~(\ref{isothermal}), this function should be a declining exponential with a scale length twice that of the light {\it if\/} the following conditions hold true:
\begin{packed_enum}
\item The uncertainties in our velocity measurements are small compared to $\sigma_z$.
\item The line-of-sight velocity residuals are dominated by the motions of stars in the $z$ direction.  
\item The scale height, $h_z$, is constant over the region of the survey.
\item The disk mass-to-light ratio, $\Upsilon$, is constant over the region of the survey.
\item The scale length, $h_R$, is constant over the region of the survey.
\end{packed_enum}

\begin{figure}[b]
\epsscale{1.2}
\plotone{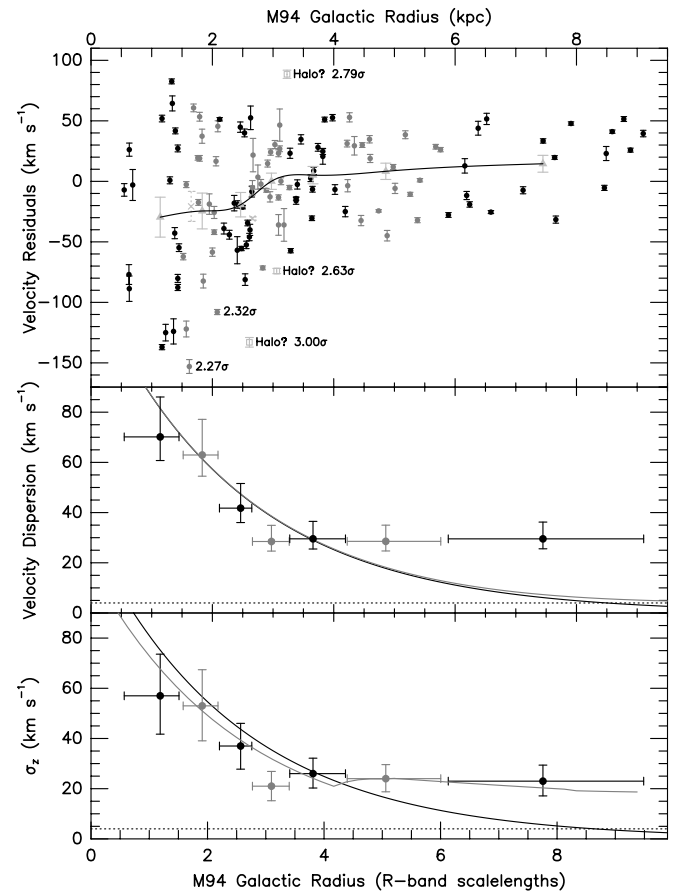}
\caption{\scriptsize M94 PN radial velocities ({\it top}), line-of-sight velocity dispersion ({\it middle}), and velocity dispersion in the $z$ direction ({\it bottom}), all as a function of galactic radius.  See the caption of Figure~\ref{ICrsds3} for more information.  Each radial bin contains 17-18 objects.  The gray line in the lower panel illustrates our best-fit two-component model for the galaxy, using an internal scale height of 300~pc.  Although the system's $z$ velocity dispersion initially falls off with the light, the profile flattens dramatically in the outer disk.  \label{M94rsds3} }
\end{figure}

\noindent
The middle panels of Figures~\ref{ICrsds3}-\ref{M101rsds3} compare our dispersions to such an exponential.  It is immediately obvious that in IC~342 and over the inner $\sim$3 scale lengths of M74 and M94, the measured velocity dispersion falls off exactly as expected for a constant $\Upsilon$ disk.  This result is consistent with that found from integrated-light absorption line spectroscopy, although the latter technique has never probed radii larger than $\sim$2~scale lengths \citep[\eg][]{b93}.  However, in the outer bins of M94, the velocity dispersion does not fall off.  Instead it asymptotes out at $\sim$30~\kms, a value that is an order of magnitude greater than the $\sim$3~\kms\ median error of the data.   This behavior is also present in the outer regions of M83:  if one ignores the low dispersions of the second and third bins (which are affected by the galaxy's outer Lindblad resonance), then the dispersion profile is consistent with a declining exponential in the inner regions, and a constant value at large radii.  Only M101 does not follow the pattern:  its line-of-sight velocity dispersion is almost flat over $\sim$4~optical scale lengths.  

Before discussing this behavior further, we consider four possible systematic errors which could be associated with our measurements.

\clearpage

\subsection{Observational Uncertainties}
\begin{figure}[t]
\epsscale{1.15}
\plotone{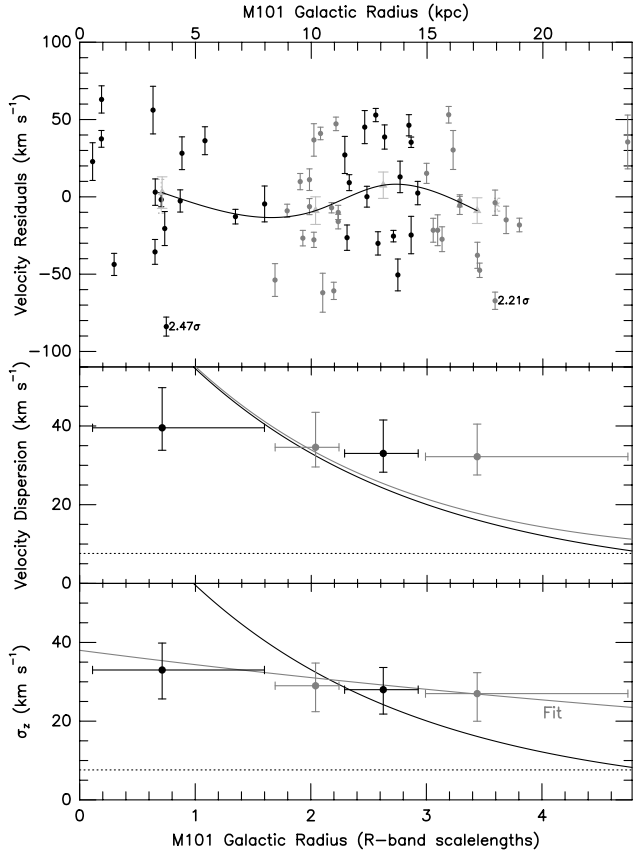}
\caption{\scriptsize M101 PN radial velocities ({\it top}), line-of-sight velocity dispersion ({\it middle}), and velocity dispersion in the $z$ direction ({\it bottom}), all as a function of galactic radius.  See the caption of Figure~\ref{ICrsds3} for more information.  Each radial bin contains 15 objects.   The gray line in the lower panel shows our best-fit exponential. Note that M101's disk velocity dispersion is roughly flat with radius. \label{M101rsds3}}
\end{figure}

Measurement errors provide a lower limit to our observed disk velocity dispersion.  If these errors were large, it could produce a plateau in the observed dispersion curve.  However, as demonstrated in Paper~II, the uncertainties quoted for our PN velocities are accurate, and significantly smaller than the measured disk dispersion (see the dotted lines in Figures~\ref{ICrsds3}-\ref{M101rsds3}).  Alternately, if some intrinsic property of the PNe themselves limited the precision of our measurements, we would again see a plateau in $\sigma_z$.  For example, if the measured radial velocity of an (asymmetrical) PN depended on its orientation, then we would expect the dispersion profile to have an artificial floor.  Of course, any such floor would need to be less than the $\sim$12~\kms\ value recorded in the outer disk of M74.  Moreover, it is difficult to conceive of how a forbidden line with a $\sim$20~\kms\ symmetric velocity width could yield a $\sim$12~\kms\ velocity error which depends on viewing angle.  In any case, since this floor should be the same in every galaxy, it cannot explain dispersions of 20 to 30~\kms\ seen in M83, M94, and M101.

One other possible error comes from the small ($\sim$5~\kms) uncertainties associated with the H{\sc i} measurements used to remove rotation, and with our estimated values for asymmetric drift.  Such an error is suppressed by $\sin i$, but it can still affect our values for $\sigma_{los}$.  To investigate this problem, we performed a series of experiments, in which the asymmetric drift was allowed to vary between 0 and 50~\kms.  These simulations showed that our line-of-sight dispersion measurements are insensitive to errors in the global rotation, changing by only a couple of \kms.  This type of error cannot be responsible for the dispersion profiles observed in the figures.

\subsection{Emission Line Contaminants}
Papers~I and II describe the criteria used to identify the planetary nebulae used in this analysis.  The data were well scrutinized, and any bias in the selection process was made on the side of eliminating valid PNe rather than including possible contaminants.  Nevertheless, it is still conceivable that compact, high-excitation H{\sc ii} regions have slipped into the kinematic sample.  The most serious case is for M101, where we have included six objects whose spectra may contain some contamination from a kinematically distinct source, and six other objects with questionably bright H$\alpha$ lines.  However, even without these objects, M101's line-of-sight velocity profile is flat with radius.

In addition, during our M74 and M83 observations, we obtained spectra of a significant number of suspected H{\sc ii} regions.  To examine the effect these objects can have on our results, we repeated the analysis of these two spirals, while including 21 and 63 contaminating sources, respectively.  As expected, the addition of a younger population decreases the observed dispersion.  In M74, the new, contaminated $\sigma_{los}$ was smaller, but within the uncertainties, still consistent with the results from the uncontaminated data.  In M83 where we added three times the number of contaminants, the dispersion again decreased, especially in the inner two bins and the outermost bin, where most of the interlopers are located.  Nevertheless, the general shape of the dispersion profile remained the same.  The results confirm that the addition of a young Population~I component can only decrease the dispersion; it cannot explain the relatively high values of $\sigma_{los}$ in the outer disks of our galaxies.

\subsection{Internal Extinction and Selection Effects}
Since spiral galaxies are known to have dust at low galactic latitudes, one might wonder if internal extinction could affect the results of our survey.  Practically speaking, a thick layer of dust would preferentially dim those PNe near the galactic plane and on the far side of the galaxy, while leaving objects with large $z$ distances on the near side unaffected.  In the case of an isothermal disk, $\sigma_z$ is independent of $z$, so dust would have no effect on our results.  However, for intermediate and exponential disks, the observed velocity dispersion increases with height above the plane \citep{vdK88}.  Consequently, a thick layer of dust could theoretically drive dispersion measurements towards higher values.  However, such a model is unlikely, as it is inconsistent with the number of PNe we observe (since the dust would extinguish large numbers of PNe below our detection limit), and the distribution of PN [\ion{O}{3}]/H$\alpha$ line ratios (see Papers~I and II).  Moreover, because the radial scale length of dust is usually larger than that for the stars \citep[][]{bianchi}, it is the dispersion measurements in the inner disk that would be most effected. \citet{M33PNe} have discussed the situation in more detail and found that dust cannot explain the flat velocity dispersions at large galactic radii.

\subsection{Incorrect Disk Models and Large Scale Warps}
Warps are known to exist in the outer regions of spiral galaxies and an undetected change in inclination could propagate into incorrect values of $\sigma_{los}$.  In fact, one of the main reasons we adjusted the geometric parameters when fitting the VLA H{\sc i} velocity maps was to remove warping as a cause for concern.  As indicated in \S4.1 and the central panels of Figure~\ref{onbands}, only M83 and M94 required anything other than a simple, flat disk.  However, since these are the best examples of flat dispersion profiles, it is important to re-examine this possibility.

M94's fit required two components, one for the galaxy's interior regions ($R < 5$~kpc), and one for the outer disk, so it is unlikely that our warping model could be an issue in this system.  The model for M83, however, was based on previous H{\sc i} analyses by \citet{M83HI} and \citet{p+01}, and was much more complicated.  Yet despite this complexity, warping only slightly alters the dispersion measurements and the overall effect on $\sigma_z(R)$ is minimal.  Specifically, if we assume a simple, flat disk, and re-run the analysis of M83's PNe, the results are essentially the same.  Even a moderate amount of warping cannot be responsible for the velocity dispersions seen in the figures.

\section{MODELING THE VELOCITY ELLIPSOID}

The velocity dispersions shown in the middle panels of Figures~\ref{M74rsds3} through \ref{M101rsds3} suggest that the outer disks of galaxies may not be simple constant scale height, constant $\Upsilon$ exponentials.  However, to confirm this result, we would like to extract $\sigma_z$ from the line-of-sight measurements.  To do this, we began by breaking down the observed velocity dispersion into its three orthogonal components: the radial ($\sigma_R$), azimuthal ($\sigma_{\phi}$) and perpendicular ($\sigma_z$) velocity dispersions, \ie
\begin{equation}
\sigma^2_{los} = \sigma^2_{\phi} \cos^2 \phi \sin^2 i + \sigma^2_R \sin^2 \phi \sin^2 i + \sigma^2_z \cos^2 i + \sigma^2_{\rm meas}.\label{siglos_eq}
\end{equation}
(Note that the (known) observational uncertainty of each velocity is explicitly included via the $\sigma_{\rm meas}$ term.)  We then removed $\sigma_{\phi}$ from the equation by taking advantage of the fact that the motion of stars in a cold disk can be approximated by the superposition of a circular orbit with movement in a small ellipse with frequency
\begin{equation}
\kappa = \frac{V_c}{R} \left( 2 + 2\frac{\partial \ln V_c}{\partial \ln R} \right)^{1/2},
\label{kappa}
\end{equation}
where $V_c$ is the circular velocity inferred from the motions of the H{\sc i} gas \citep{bt87}.   The azimuthal and radial velocity dispersions are therefore related by
\begin{equation}
\sigma^2_{\phi} = \sigma^2_R \left( \frac{1}{2} + \frac{1}{2} \frac{\partial \ln V_c}{\partial \ln R} \right).
\label{epicyclic}
\end{equation}
The line-of-sight dispersion can then be written as a function of two variables: $\sigma_R$, which modulates with position angle in the galaxy, and $\sigma_z$, which does not.  We can therefore determine the likelihood that a particular combination of these two variables will generate an observed set of PN velocities.

Figures~\ref{ICcon}-\ref{M83con} display the relative probabilities of such an analysis.  For the calculation, we have constrained the ratio of the two dispersions to lie in the range $0.25 < \sigma_z / \sigma_R < 1.0$, where the upper limit comes from the physics of disk scattering, since both theory \citep{v85, jb90} and observations \citep{b99, g+97, g+00} require $\sigma_R$ to dominate.  The lower limit is a consequence of the buckling or ``fire hose'' instability, which can occur in a cold disk \citep{t66, a85, ms94}.  These limits are responsible for the cone-like shapes of the diagrams.

As the figures show, the shapes of the likelihood probabilities vary from galaxy to galaxy.  This reflects differences in galactic inclination.  In M74, which is extremely close to face-on ($i \sim 6^{\circ}$), $\sigma_{los}$ is almost entirely due to $\sigma_z$, and $\sigma_R$ is essentially unconstrained.  Conversely, as the disks become more inclined, our sensitivity to $\sigma_R$ grows at the expense of our measurement of $\sigma_z$.

The bottom panels of Figures~\ref{ICrsds3}-\ref{M101rsds3} marginalize these likelihoods over radial dispersion to find the most likely values of $\sigma_z$ and their $1\,\sigma$ uncertainties.  Once again, the black curves indicate exponential laws with twice the scale length of the $R$-band light. As expected for low-inclination systems, the $\sigma_z$ profiles track those of the line-of-sight dispersions extremely well.   The ratio of $\sigma_z$ to $\sigma_{los}$ is roughly constant across the face of each galaxy, ranging from $\sim$0.76 in IC~342 to $\sim$0.86 in M74.  For IC~342 and M74, the disks are well-fit by a constant scale height, constant $\Upsilon$ model.  In contrast, the data for M83 and M101 do not follow the expected exponential:  in these objects, either the scale length, scale height, or mass-to-light ratio is changing.  Finally, M94 appears to be a hybrid: while the inner $\sim$4 scale lengths can be fit with a simple model, the outer regions demand an additional term.

One additional constraint can be applied to the data.  In order for our thin stellar disks to be stable against axisymmetric perturbations, they must satisfy the \citet{t64} criterion
\begin{equation}
\sigma_R > \frac{3.36 G \Sigma}{\kappa}.
\label{Toomre}
\end{equation}
If we combine this relation with equation~(\ref{isothermal}), we obtain a new constraint on the possible values of $\sigma_R$ and $\sigma_z$
\begin{equation}
\sigma_z < \left( \frac{K h_z \kappa \sigma_R}{3.36} \right)^{1/2} \label{ToomStab}
\end{equation}
where $K = 1.7051\pi$ for intermediate disks, $h_z$ is the stellar scale height, and $\kappa$ is the epicyclic frequency given by equation~(\ref{kappa}). 

\begin{deluxetable*}{ccccccccc}
\tabletypesize{\scriptsize}
\tablecaption{Disk Mass Models \label{massmodel}}
\tablewidth{0pt}
\tablehead{
\colhead{Galaxy}&\colhead{Filter}&\colhead{$\mu(0)$}&\colhead{$h_z$ (pc)}&\colhead{$h_R$ (kpc)}&\colhead{$\sigma_z(0)$ (\kms)} &\colhead{$\Sigma(0) (M_{\odot}~{\rm pc}^{-2}$)} &\colhead{$\Upsilon(0)$} &\colhead{$d\log(\Upsilon h_z) /dr$}
}
\startdata
IC 342  &$I$   &$19.60 \pm 0.20$ &500 &4.24 &$41 \pm 6$         &$146 \pm 45$                 &$0.25 \pm 0.10$ &0 \\
M74      &$R$ &$20.20 \pm 0.05$ &400 &3.17 &$53 \pm 5$         &$305^{+65}_{-48}$        &$1.4 \pm 0.3$ &0 \\
M83      &$R$ &$19.07 \pm 0.05$ &400  &4.0   &73                        &578                                   &$1.0 \pm 0.5$  &0         \\
              &\dots&\dots                        &1200 &10   &40                        &\dots                                 &\dots                  &0.16  \\  
M94      &$R$ &$18.88 \pm 0.05$ &300 &1.22 &$91 \pm 15$       &$1200^{+460}_{-370}$ &$1.8 \pm 0.4$ &0 \\
              &$R$ &$22.69 \pm 0.10$ &900 &13.2 &$29 \pm 5$         &$41 \pm 15$                    &$2.0 \pm 0.4$ &0 \\
M101   &$R$  &$20.29 \pm 0.10$ &600  &24.8 &$38^{+8}_{-4}$ &$100^{+50}_{-20}$         &$0.6 \pm 0.3$ &0.07
\enddata
\end{deluxetable*}

The solid curves in Figures~\ref{ICcon}-\ref{M83con} show the maximum values of $\sigma_z$ allowed by the Toomre criterion, given several different estimates of $h_z$ (see Table~\ref{tabhz}).  For example, in IC~342, the assumed range of $h_z$ is consistent with the derived likelihood probabilities; if we were to hypothesize a much lower value for $h_z$, then the likelihood of a stable solution would decrease dramatically.  Similarly, for most of M74 and M101, the probabilities lie safely below the Toomre curves in the region of stability.  However, in the galaxies' outermost regions, slightly larger scale heights may be in order.  Finally, the probabilities for M83 and M94 suggest that our assumed values of $h_z$ are much too small, especially at large galactic radii.   This result may not be too surprising, since these are also the objects that have the most noticeable asymmetric drift.  It is likely that the energy missing from disk rotation has, at least in part, been translated into larger $z$ motions for the stars, and a greater galactic scale height.

\section{INDIVIDUAL GALAXIES}

Table~\ref{massmodel} summarizes our results for the five galaxies in our program.  Each object is discussed below.

\subsection{IC~342}
IC~342 is a large ($R > 20$~kpc), nearby ($D \sim 3.5$~Mpc) Scd spiral that overfills the $\sim$10$\arcmin \times 10\arcmin$ field-of-view of the WIYN OPTIC imager.  Consequently, our survey was limited to the galaxy's inner $\sim$2 scale lengths, a regime previously probed (in other, more distant galaxies) with integrated-light absorption line spectroscopy \citep[\eg][]{b93, g+97, g+00}.  Since these studies have generally found that the mass-to-light ratio of inner spiral disks is constant with radius, it is not surprising that we find a similar result.  

\begin{figure}
\epsscale{1.22}
\plotone{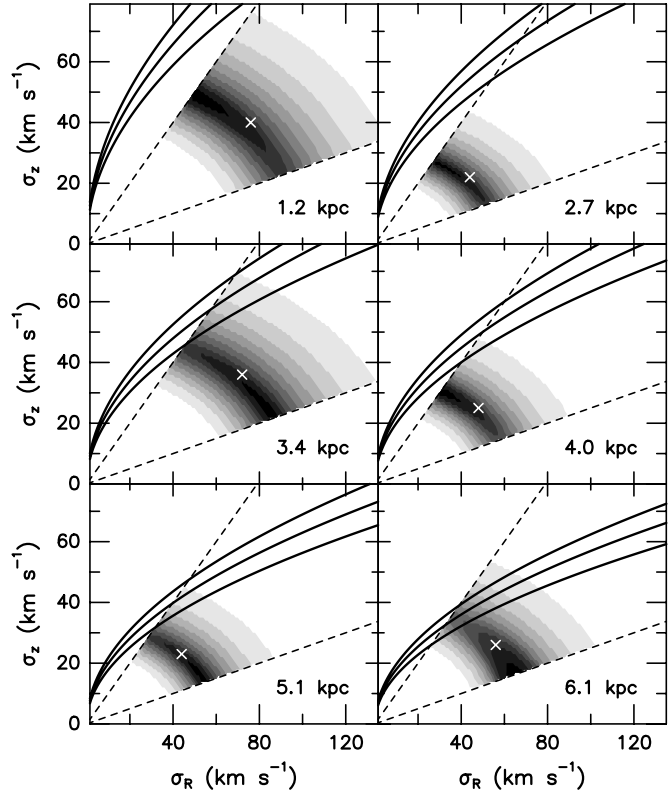}
\caption{\scriptsize Likelihood probabilities for IC~342, as a function of $\sigma_z$ ($y$-axis) and $\sigma_R$ ($x$-axis).  The darkness is directly proportional to probability, and the white crosses indicate our most likely solutions, without any consideration for stability.  The solid curves illustrate the maximum value of $\sigma_z$, above which the disk becomes unstable to axisymmetric perturbations \citep{t64}, and correspond to disk scale heights of $h_z$ = 400, 500, and 600~pc (from bottom to top). The figure illustrates that our estimates of $h_z$ are reasonable, while smaller values are problematic, due to stability concerns.  \label{ICcon} }
\end{figure}

The exponential curve shown in Figure~\ref{ICrsds3} is derived from the $B$, $V$, and $I$ surface photometry of \citet{bm99} and \citet{wb03}, which shows that, over the region of our survey, the disk of IC~342 is well fit by a single exponential with a scale length of $250\arcsec$ ($\sim$4.2~kpc).  (There is a possible steepening of the profile at $R \gtrsim 700\arcsec$ but that is well beyond our frame limit.)   Our inferred values of $\sigma_z$ follow this curve very well, yielding a dynamical scale length ($\sim$5~kpc) that is not significantly larger than that of the $I$-band light.  Moreover, all our values for $\sigma_z$ lie safely under the \citet{t64} stability curves of Figure~\ref{ICcon}, suggesting that our estimate of $h_z$ is not unreasonable.

Our most-likely value for the central $z$-direction velocity dispersion of IC~342 is $\sigma_z(0) = 41 \pm 6$~\kms.  If we assume that the disk $\Upsilon$ is constant, and adopt a scale height of $h_z = 500$~pc, then through equation~(\ref{isothermal}), the galaxy's central disk surface mass density is $\Sigma(0) = 146 \pm 45 \, (h_z / 500~{\rm pc})^{-1} \, M_{\odot}$~pc$^{-2}$.  A comparison of this number with the galaxy's central disk surface brightness \citep[$\mu_I(0) = 19.6 \pm 0.2$;][]{wb03} then yields an $I$-band disk mass-to-light ratio of $\Upsilon_I \sim 0.25 \pm 0.10 \, (h_z / 500~{\rm pc})^{-1}$ for $E(B-V) = 0.558 \pm 0.069$ \citep{sfd98} and a \citet{ccm89} reddening law.  Note that our error includes the uncertainties in the fit, the surface photometry, and $h_z$ (estimated at $\sim$20\%), but not the systematic error associated with the disk's thermal structure.  As described in \S3, this error can be as high as $\sim$30\%, but it is most likely a systematic offset that affects all our measurements.  

\subsection{M74 (NGC~628)}
M74 is the most face-on galaxy in our sample, and the observed line-of-sight velocity dispersion is almost entirely due to $\sigma_z$.  Like IC~342, multi-band surface photometry confirms that the disk is well fit by a single exponential over the entire range of our survey \citep{npr92, khc06}.  The galaxy, does, however, have a slightly earlier Hubble type (Sc) than IC~342, and is slightly smaller:  at Paper~I's distance of 8.6~Mpc, M74's $R$-band disk scale length ($\sim$3.2~kpc) is $\sim$25\% shorter than that of the former galaxy \citep{m04}.  M74's maximum rotation velocity is also smaller than that of IC~342.

\begin{figure}
\epsscale{1.21}
\plotone{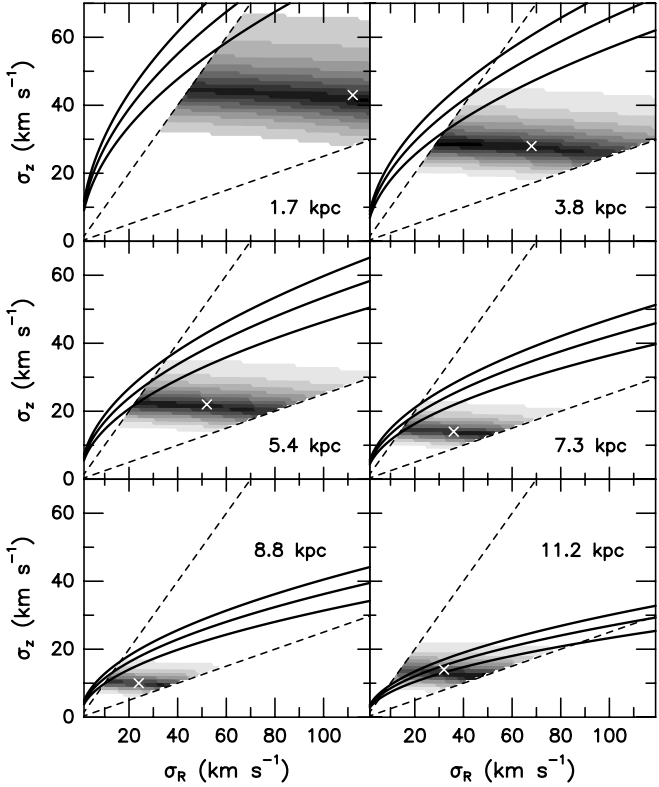}
\caption{\scriptsize Likelihood probabilities for M74, as a function of $\sigma_z$ ($y$-axis) and $\sigma_R$ ($x$-axis).  See the caption of Figure~\ref{ICcon} for more information.  The three curves show the stability limits for $h_z$ = 300, 400, and 500~pc (from bottom to top); solutions above these curves are excluded by the \citet{t64} condition.  Within $\sim$10~kpc, a constant scale height of $\sim$400~pc is reasonable, but for the final bin, a larger value of $h_z$ may be needed. \label{M74con} }
\end{figure}

As seen in Figure~\ref{M74rsds3}, a constant $h_z$, constant $\Upsilon$ model is an excellent fit to M74's velocity dispersion.   The most likely scale length for $\sigma_z^2$, $3.3^{+1.1}_{-0.5}$~kpc, is statistically indistinguishable from that derived from the $R$-band light, and, with the possible exception of the outermost bin, the probabilities of Figure~\ref{M74con} are all well below the \citet{t64} stability curve.  In addition, the disk's central velocity dispersion, as extrapolated from the $\sigma_z$ profile ($53 \pm 5$~\kms) is in good agreement with the $\sim$60~\kms\ value found from integral-field absorption-line spectroscopy of the system's central kiloparsec \citep{g+06}.  Our measurement, coupled with our adopted scale height estimate of $h_z = 400$~pc, implies a central surface mass of $\Sigma(0) = 305^{+65}_{-48} \, (h_z / 400~{\rm pc})^{-1} \, M_{\odot}$~pc$^{-2}$.  The surface photometry of \citet{m04} combined with a foreground extinction of $E(B-V) = 0.07$ \citep{sfd98} then yields $\Upsilon_R \sim 1.4 \pm 0.3 \, (h_z / 400~{\rm pc})^{-1}$, or $\Upsilon_I \sim 1.0 \pm 0.3 \, (h_z / 400~{\rm pc})^{-1}$.  The fact that this mass-to-light ratio is larger than that of IC~342 is consistent with the galaxy's earlier Hubble type.

\subsection{M101 (NGC~5457)}
\begin{figure}[b]
\epsscale{1.22}
\plotone{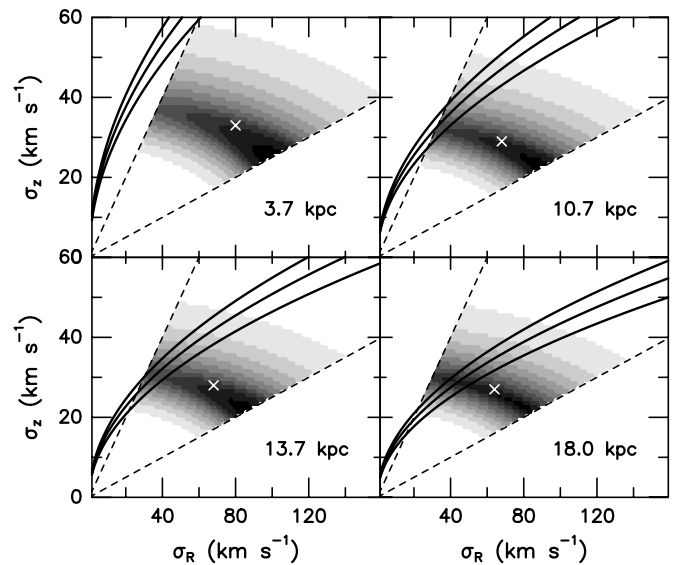}
\caption{\scriptsize Likelihood probabilities for M101, as a function of $\sigma_z$ ($y$-axis) and $\sigma_R$ ($x$-axis).  See the caption of Figure~\ref{ICcon} for more information.  The three curves show the stability limits for $h_z$ = 500, 600, and 700~pc (from bottom to top).  As with M74, the data for the inner bins are consistent with the nominal scale height, but in the outer disk, a larger scale height might be needed. \label{M101con} }
\end{figure}

M101, a grand design Scd spiral, is the largest galaxy in our sample, with an $R$-band scale length of $\sim$5~kpc \citep{khc06}, and a maximum rotation speed of $\sim$250~\kms.  Unfortunately, it is also the galaxy with the poorest velocity sampling, with only 60 PN measurements, including 12 objects whose velocities may be compromised by nearby emission.  Nevertheless, since the system's inclination is only $\sim$17$^\circ$, the probabilities produced by our analysis (Figure~\ref{M101con}) are relatively flat, and the corresponding values of $\sigma_z$ are reasonably well-defined.  Interestingly, the dispersion profile derived from our measurements looks nothing like that of a constant $h_z$, constant $\Upsilon$ disk.  As Figure~\ref{M101rsds3} illustrates, the best-fitting scale length for $\sigma_z^2$ is $\sim$5~times larger than that of the galaxy's luminosity.

To explain this result, we first consider the possibility that the scale height, $h_z$, is changing over the area of our survey.  Figure~\ref{M101con} presents no evidence for this either way, as the probabilities derived from the PN velocities are generally below the constraint imposed by the \citet{t64} condition.   However, in order to mimic a constant $\Upsilon$ disk, $h_z$ would have to increase by a full order of magnitude between $\sim$3.5~kpc and $\sim$18~kpc.  Such a gradient is unlikely, as it would make M101 unique:  none of the large, edge-on spirals observed by \citet{vdKs82} and \citet{bm02} have such a profile.

Alternatively, we can assume that $h_z$ is constant, and ascribe the flat velocity dispersion to a change in the galaxy's disk mass-to-light ratio.  M101 has a central $R$-band surface brightness of $\mu_R(0) \sim 20.29$ \citep{k+00}, and almost no foreground extinction \citep[$E(B-V) = 0.009$;][]{sfd98}.  These numbers, combined with the most-likely value for the disk's central velocity dispersion, $\sigma_z(0) = 38^{+8}_{-4}$~\kms, imply a central disk surface mass of $\Sigma(0) = 100^{+50}_{-20} \, (h_z / 600~{\rm pc})^{-1} \, M_{\odot}$~pc$^{-2}$, and a central disk mass-to-light ratio of $\Upsilon_R \sim 0.6 \pm 0.3 \, (h_z / 600~{\rm pc})^{-1}$.  These values are reasonable:  the surface mass is consistent with that inferred from the H{\sc i} rotation curve \citep{THINGS}, and $\Upsilon$ falls midway between that of the blue, late-type spiral IC~342, and that for the earlier M74 system.  However, when we adopt these numbers and use the observed $R$-band disk scale length of $\sim$5~kpc, the velocity dispersion at $\sim$17~kpc implies a mass-to-light ratio of $\Upsilon_R \sim 9 \, (h_z / 600~{\rm pc})^{-1}$.  This is an extremely large value for an actively star-forming disk.

Again, we note that due to the limited number of planetary nebulae discovered in the \citet{M101PNe} survey, the velocity field of M101 is poorly sampled.  With only four bins of 15~PNe each, there is not enough information to model the galaxy with both a variable $h_z$ and changing $\Upsilon$.  Moreover, with such a limited number of objects, our analysis is susceptible to the effects of disk substructure.  For example, if some of the PNe targeted in our survey actually belong to the remnants of a disrupted companion galaxy (like the Milky Way's Sagittarius dwarf), then our derived disk mass measurements would be in error.  The only definite statement we can make about M101 is that further observations are warranted.

\subsection{M94 (NGC~4736)}
The Sab spiral M94 is the earliest galaxy in our sample, with a small, but high surface-brightness bulge that inhibits PN detections in the system's central $\sim$0.5~kpc.  Nevertheless, because the galaxy is relatively small \citep[$h_R = 1.22$~kpc;][]{g+09,dB+08}, nearby ($\sim$4.4~Mpc), and has little foreground extinction \citep[$E(B-V) = 0.018$;][]{sfd98}, it is the system for which we have the best data (median PN velocity uncertainty of $\sim$3~\kms) with a wide spatial sampling (out to $\sim$7 disk scale lengths).  The result is the high-quality dispersion profile seen in Figure~\ref{M94rsds3}, which clearly shows an exponential decline over the galaxy's inner regions, followed by a flattening which begins at $\sim$4~disk scale lengths.

Unlike the galaxies of the previous sections, M94 does not have a simple surface brightness profile:  while the galaxy's inner disk follows an exponential law, its outer regions are dominated by the luminosity of a faint stellar ring \citep{s61, ss76}.  \Citet{dB+08} have recently analyzed M94's profile using 3.6~$\mu$m surface brightness measurements from the {\sl Spitzer\/} Infrared Nearby Galaxy Survey \citep[SINGS;][]{SINGS}.   Their data show that the galaxy's disk is best modeled by an inner, rapidly falling exponential (scale length $h_{R,i} = 1.22$~kpc), which is abruptly replaced at $R \sim 5$~kpc by an almost constant surface-brightness (scale length $h_{R,o} = 7.16$~kpc) envelope.   The surface photometry of \citet{g+09} confirms this result:  in the $R$-band, M94 has a rapidly declining inner disk  ($h_{R,i} = 1.22$~kpc), a uniform brightness outer ring ($h_{R,o} = 13.2$~kpc between 4.6 and 7.0~kpc), and a slowly declining outer disk ($h_{R,x} = 2.98$~kpc).  Thus,  in the classification scheme of \citet{e+05}, M94 is a Type~III ``antitruncated'' spiral, whose disk surface brightness declines with one slope in its inner regions, and a different, shallower slope at larger radii.   Roughly 30\% of all spirals, and $\sim$50\% of all early-type spirals fall into this category \citep{p+08}.  Furthermore, our PN number counts support the classification of M94 as a Type~III object, as there are far too many planetary nebulae at large radii to be explained by a single exponential law.
  
\begin{figure}[b]
\epsscale{1.22}
\plotone{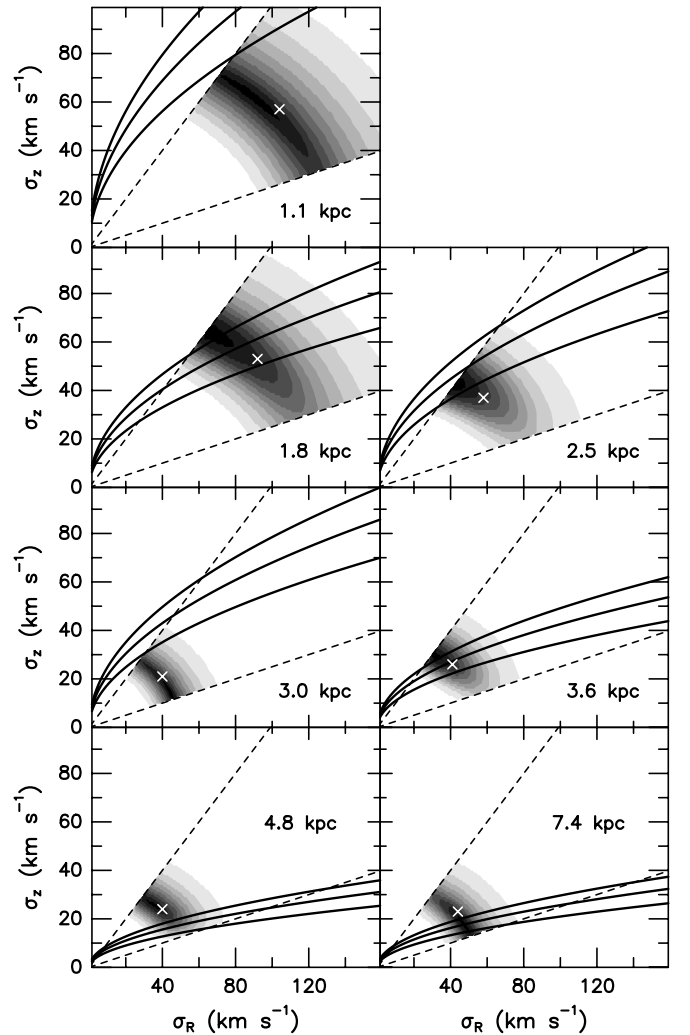}
\caption{\scriptsize Likelihood probabilities for M94, as a function of $\sigma_z$ ($y$-axis) and $\sigma_R$ ($x$-axis).  See the caption of Figure~\ref{ICcon} for more information.  The three curves show the stability limits for $h_z$ = 100, 200, and 300~pc (from bottom to top).  In the inner five bins, a scale height of $h_z$ = 300~pc appears consistent with the data, but at larger radii, a much larger scale height is required. \label{M94con} }
\end{figure}

While the cause of the break in Type~III disk galaxies is unclear, one possibility is that it is due to the increased importance of the thick disk at large radii \citep{yd06, p+07}.  Our kinematic study is consistent with this hypothesis.  Since early-type spirals typically have $h_R / h_z \simlt 10$ \citep{bm02, dG98, kvd02}, M94's disk scale height should be relatively large, $h_z \sim 300$~pc. Indeed, as Figure~\ref{M94con} illustrates, a 300~pc scale height satisfies the \citet{t64} stability criterion throughout the inner regions of the galaxy, while values much smaller than $\sim$200~pc do not.  However, outside $\sim$4~kpc, no thin-disk value of $h_z$ is stable; at these radii, significantly larger scale heights, even as high as 900~pc, are needed.  In M94's outer regions, the light we see must be coming from a thick disk.   

Based on these data, we decided to model the galaxy using two separate kinematic populations.  Within 3.8~kpc, we assume that most of the PNe are part of the galaxy's thin disk with $h_z \sim 300$~pc; outside of 5~kpc, we allow the thick disk to dominate, and set the scale height to 3~times that of the thin disk, \ie\ 900~pc.  Such a ratio agrees with observations of the Milky Way \citep{lh03} and NGC~891 \citep{m+97}, and is consistent with the range of values found by \citet{yd06} for a sample of 34 edge-on galaxies.  In between these two extremes, we define a transition zone, where the effective scale height increases linearly with radius.

The gray curve in the lower panel of Figure~\ref{M94rsds3} illustrates our thin disk-thick disk model, with each component having a fixed mass-to-light ratio.   This curve is an excellent fit to the data.  The central disk velocity dispersion implied by our fit is $91 \pm 15$~\kms; when combined with M94's central $R$-band disk surface brightness \citep[$\mu_R(0) = 18.88 \pm 0.05$;][]{g+09} and foreground extinction \citep[$E(B-V) = 0.018$;][]{sfd98}, this yields a thin-disk mass-to-light ratio of $\Upsilon_{R} = 1.8 \pm 0.4 \, (h_z / 300~{\rm pc})^{-1}$.  Our thick-disk mass-to-light ratio based on this same $R$-band surface photometry and the outer three bins of Figure~\ref{M94rsds3} is $\Upsilon_{R} = 2.0 \pm 0.4 \, (h_z / 900~{\rm pc})^{-1}$.  These values are reasonable, given the galaxy's early Hubble type.

\subsection{M83 (NGC~5236)}
The barred spiral M83 is the galaxy for which we have the widest spatial coverage ($\sim$25~kpc) and the largest set of high-precision PN velocities (162).  Unfortunately, it is also the object with the most puzzling dispersion profile.  As Figure~\ref{M83rsds3} illustrates, except for its innermost bin, M83's $\sigma_z$ values remain virtually unchanged from $\sim$6 to $\sim$20~kpc, a region which spans $\sim$4~disk scale lengths.  One reason for this behavior may be the presence of the system's strong bar, as the structure's outer Lindblad resonance is located between our second and third velocity bins \citep[at $290\arcsec$ or $\sim$6.83~kpc;][]{L+04}.  Since all the stellar velocities in the area are affected, it is quite possible that our derived value for the region's $\sigma_z$ is an underestimate.  These points should therefore be excluded from the analysis.

\begin{figure}
\epsscale{1.21}
\plotone{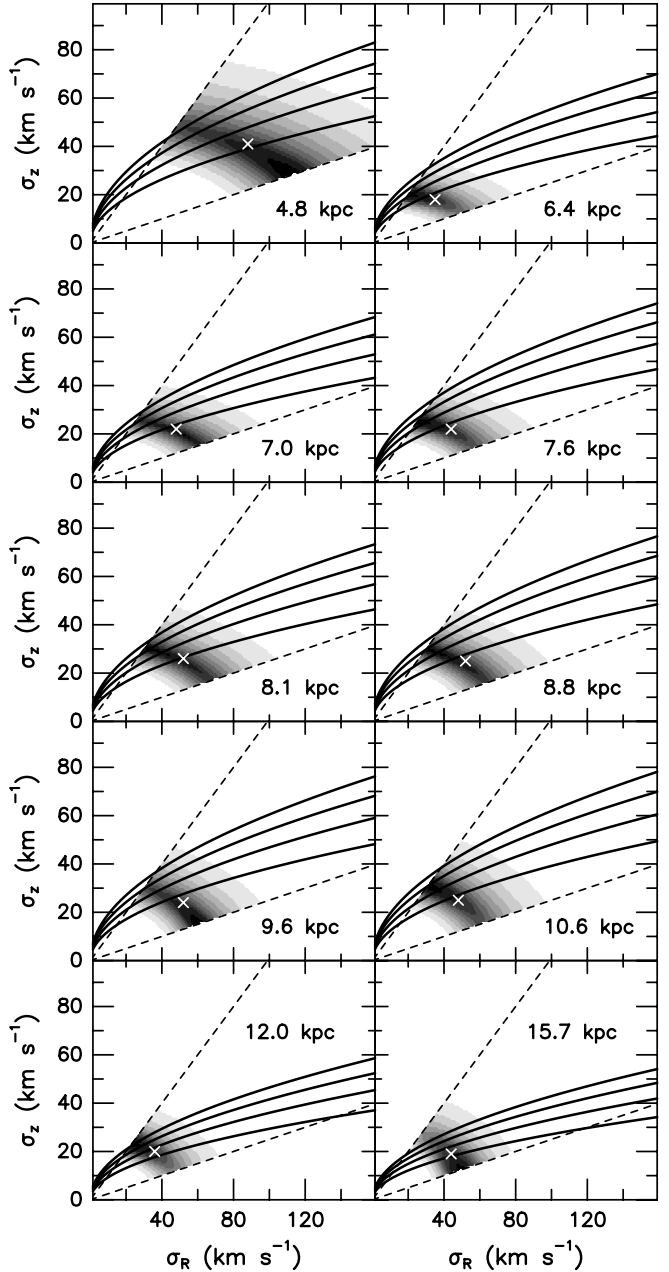}
\caption{\scriptsize Likelihood probabilities for M83, as a function of $\sigma_z$ ($y$-axis) and $\sigma_R$ ($x$-axis).  See the caption of Figure~\ref{ICcon} for more information.  The four curves show the stability limits for $h_z$ = 200, 300, 400, and 500~pc (from bottom to top).  As in M94, the data suggest that larger scale heights are needed to maintain stability against axisymmetric perturbations \citep{t64}. \label{M83con} }
\end{figure}

The rest of M83's dispersion profile is not as easy to interpret.  Unlike M94, M83 shows no evidence for a Type~III anti-truncated disk or a flattened surface-brightness profile at large radii.  Indeed, according to de Jong (2008, private communication), the galaxy's $H$-band surface brightness profile declines with a single exponential scale length of 80$\arcsec$ (1.88~kpc) over the entire region of our survey.   Thus, the relatively large velocity dispersion in the outer disk cannot be explained by an excess of stellar mass in the region.

Another puzzle comes from the galaxy's rather small radial scale length.  Although M83 is the fastest rotator in our sample (with a $v_{max}$ of 255 \kms), it also has the second smallest value of $h_R$: 2.45~kpc in $R$ and 1.88~kpc in $H$ \citep[][de Jong 2008, private communication]{k+00}.  If the scale length to scale height ratio is similar to that of other galaxies, \ie\ $h_R / h_z \sim 10$ \citep{bm02, dG98, kvd02}, then this implies a value for $h_z$ near $\sim$200~pc.  However, as Figure~\ref{M83con} demonstrates, scale heights as small as this yield solutions for $\sigma_z$ that are unstable to axisymmetric perturbations \citep{t64}.  As a result, any attempt to fit the data by varying $h_z$ while keeping $h_R$ constant leads to either an unphysically large scale height, an exceptionally low value of $h_R/h_z$, or a large central velocity dispersion that is completely inconsistent with the galaxy's rotation curve.

Given these difficulties, we chose to model M83 with a two component, thin disk-thick disk model.  For the galaxy's thin disk, we adopted a scale length of 4.0~kpc, a central velocity dispersion of $\sigma_z(0) \sim 73$~\kms, and a scale height of 400~pc; for the systems' thick disk, we assumed $h_R \sim 10$~kpc, $\sigma_z(0) = 40$~\kms, and $h_z = 1200$~pc. As the curve in the lower panel of Figure~\ref{M83rsds3} indicates, this type of model, although arbitrary, does do a good job of reproducing the observed behavior of $\sigma_z$.

If we adopt this model, and use a central $R$-band disk surface brightness of $\mu(R) = 19.07$ \citep{k+00} and a foreground Galactic extinction of $E(B-V) = 0.066$ \citep{sfd98}, then M83's central disk mass-to-light ratio becomes $\Upsilon_R = 1.0 \pm 0.5 \, (h_z / 400~{\rm pc})^{-1}$, where the 50\% uncertainty is a conservative estimate.   In keeping with the galaxy's Hubble type, this value is intermediate between the lower values of IC~342 and M101 and the high values for M74 and M94.  Of course, because the galaxy's surface brightness declines with radius while its $z$ velocity dispersion does not, the inferred value for $\Upsilon$ increases dramatically in the galaxy's outer regions.

\section{DISCUSSION}

Except for M101, the data from Figures~\ref{ICrsds3} through \ref{M101rsds3} offer a consistent picture for the mass profile of spiral disks.  Within the inner $\sim$3 to 4 disk scale lengths, the ratio of $\sigma_z^2 / \mu(R)$ in IC~342, M74, and M94 is flat with radius.  Through equation (\ref{isothermal}), this implies that the stellar vertical velocity dispersion is consistent with that expected from a constant $h_z$, constant $\Upsilon$ disk.  (The data for M83 can also be interpreted in this way, if one accepts that the stellar kinematics between $\sim$6 and $\sim$7~kpc are affected by the bar's outer Lindblad resonance).  In the latest-type spirals, IC~342 and M101, the disk mass-to-light ratio is less than one; in the earlier systems, $\Upsilon_R$ is roughly between one and two.   Although these numbers have a degree of systematic uncertainty due to our lack of knowledge about the disks' vertical structure, they are nonetheless in the range expected for normal, star-forming stellar populations.

\begin{figure}[t]
\epsscale{1.22}
\plotone{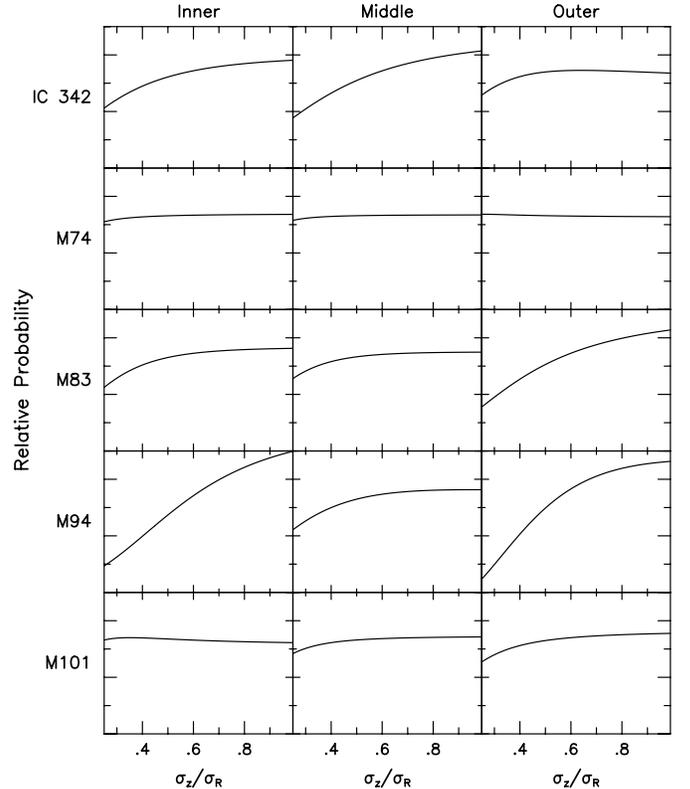}
\caption{\scriptsize Relative likelihoods for $\sigma_z/\sigma_R$ in the five disks observed in this survey.  The PNe in each galaxy were divided as equally as possible into three regions (inner, middle, and outer).  Because $\sigma_R$ is poorly constrained in these low-inclination systems, the likelihood curves are all very broad.  In our moderately inclined galaxies (IC~342, M83, and M94), the results seem to favor higher values of $\sigma_z/\sigma_R$, but this could be an artifact caused by switching from $\sigma_R$-$\sigma_z$ space to ($\sigma_z/\sigma_R$)-$\sigma_z$ space.  There is no evidence for a shape gradient in any of the galaxies. \label{shapes} }
\end{figure}

Our measurements of the disk velocity ellipsoids are quite uncertain.  Since the galaxies in our sample were selected on the basis of their low inclination, our ability to measure $\sigma_R$, the disk velocity dispersion in the radial direction, is very limited.  Consequently, as Figure~\ref{shapes} illustrates, a marginalization over the likelihoods of Figures~\ref{ICcon}-\ref{M83con} places few constraints on $\sigma_z/\sigma_R$, especially in our two most face-on galaxies, M74 and M101.  The trend toward higher $\sigma_z/\sigma_R$ in the three more inclined systems is likely a numerical artifact caused by switching from $\sigma_R$-$\sigma_z$ space to ($\sigma_z/\sigma_R$)-$\sigma_z$ space for this analysis.


Finally, the data in Figure~\ref{shapes} can be used to confirm our estimates of each galaxy's asymmetric drift.   The full expression for the drift velocity in a stellar disk is
\begin{eqnarray}
V_c^2 - \langle V_{\ast} \rangle^2 &=& \sigma_R^2 \left( \frac{\sigma_{\phi}^2}{\sigma_R^2} - 2 \frac{\partial \ln \sigma_R}{\partial \ln R} - \frac{\partial \ln \nu}{\partial \ln R} - 1 \right) \nonumber \\
& &- R\frac{\partial \sigma_{Rz}}{\partial z} - \frac{R}{\nu} \frac{\partial \nu}{\partial z} \sigma_{Rz},
\label{Jeans0}
\end{eqnarray}
where $\nu$ is the stellar density \citep{bt87}.  However, if the disk is roughly isothermal with constant $h_z$, then the epicyclic approximation gives
\begin{eqnarray}
V_c^2 - \langle V_{\ast} \rangle^2 &\simeq& 2 V_c v_{asd} \\
 &\simeq& \sigma_R^2 \left( \frac{1}{2} \frac{\partial \ln V_c}{\partial \ln R} - 2 \frac{\partial \ln \sigma_R}{\partial \ln R} - 2 \frac{\partial \ln \sigma_z}{\partial \ln R} - \frac{1}{2} \right). \nonumber
\label{Jeans1}
\end{eqnarray}
Finally, Figure~\ref{shapes} suggests that the shape of a disk's velocity ellipsoid does not change much with radius.  In this case, the application of equation~(\ref{isothermal}) creates a very simple expression for asymmetric drift
\begin{equation}
v_{asd} \simeq {\sigma_R^2 \over V_c} \left(\frac{1}{4} \frac{\partial \ln V_c}
{\partial \ln R} + \frac{R}{h_R} - \frac{1}{4} \right)
\label{Jeans2}
\end{equation}
where we have substituted in $h_R$ by assuming a constant mass-to-light ratio in the stellar disk.  

This equation yields roughly constant asymmetric drift velocities throughout the three disks where our assumptions appear valid.  In the inner regions of our earliest spiral, M94, equation~(\ref{Jeans2}) implies an asymmetric drift velocity of $v_{asd} \sim 40$~\kms,  exactly the value assumed in \S 4.2.  Conversely, in the two latest galaxies in our sample, our derived values for the drift velocity, $v_{asd} \sim 11$~\kms\ (for IC~342) and $\sim$8~\kms\ (for M101) are slightly lower than the values assumed above (20 and 25~\kms, respectively).  Of course, given the uncertainties associated with our estimates of $\sigma_R$, the numbers are still consistent.  Finally, the face-on nature of M74 and disturbed rotation curve of M83 (due to its bar), preclude statements about the asymmetric drifts in these galaxies.  However, it does appear that our assumption of asymmetric drift being between 10\% and 20\% of the rotation speed is valid.

\subsection{Inner Disks}

\begin{figure*}
\epsscale{1.15}
\plotone{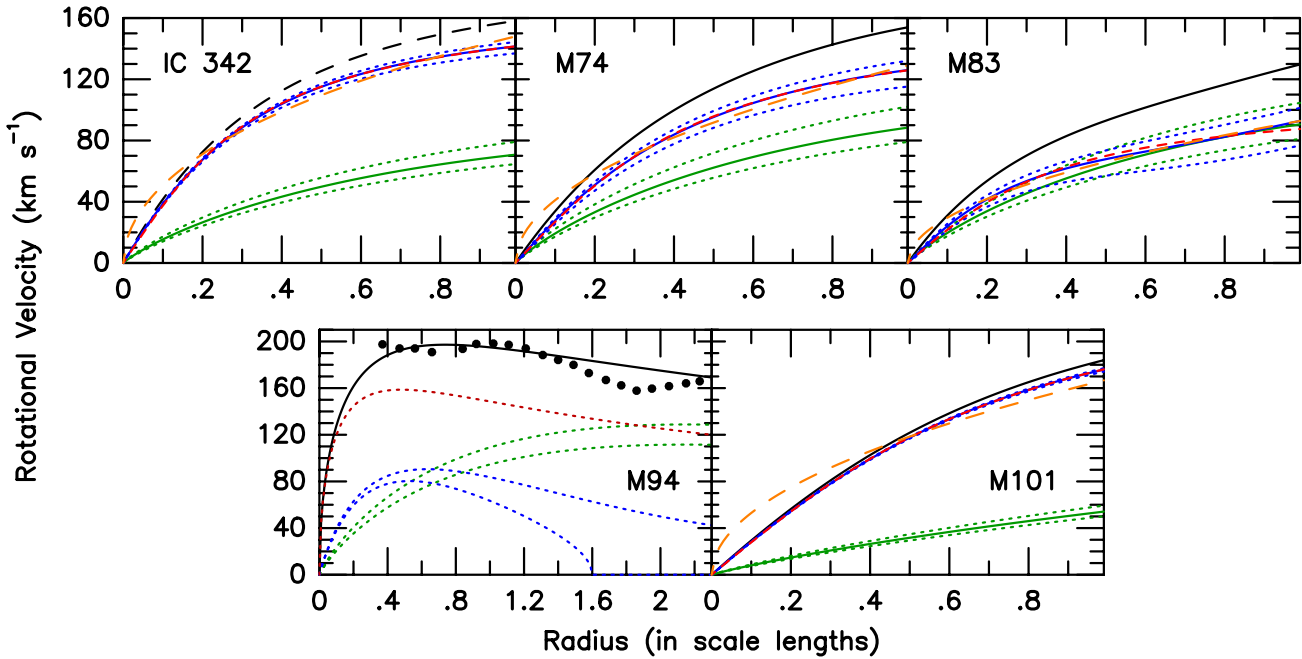}
\caption{\scriptsize The rotation curves for the inner disks of our five program galaxies. The solid black curves display the general motion of the H{\sc i} gas (the black dots in M94 show the THINGS rotation curve from \citet{dB+08}); the green curves show the contribution of the disk under the assumption of an intermediate (sech$(z / h_z)$) vertical structure and the values of $h_z$ given in the last column of Table~\ref{tabhz}.  The blue curves show the contribution of the residual halo component, with the NFW profiles shown with long orange dashes and the pseudo-isothermal models denoted by short red dashes.  The bulge of M94 (indicated by a dotted dark red curve) is a substantial and uncertain contributor to the mass, so no residual halo was fit to this system.  (See the on-line article for a color version of this figure.) \label{rotcurves} }
\end{figure*}

Most analyses of the rotation curves of spirals galaxies make use of the maximal-disk hypothesis, wherein the rotation curve near the center of the galaxy is assumed to be defined entirely by the potential of a thin disk.  Our measurements of $\sigma_z$ allow us to test this assumption and probe the distribution of the systems' dark matter halos.  The circular velocity produced by an exponential disk with radial scale length, $h_R$, and central surface mass, $\Sigma_0$, is given by
\begin{equation}
V_c^2(R) = 4\pi G \, \Sigma_0 h_R \, y^2 [I_0(y)K_0(y) - I_1(y)K_1(y)]
\label{disk_vc0}
\end{equation}
where $y = R/(2h_R)$ and $I_n$ and $K_n$ are modified Bessel functions \citep{f70, bt87}.  When combined with equation~(\ref{isothermal}), this implies a relation between the disk's rotation speed, $z$-velocity dispersion, and scale length to scale height ratio, \ie
\begin{equation}
\frac{V_c^2(R)}{\sigma_z^2(R)} = \frac{4\pi}{K} \left( \frac{h_R}{h_z} \right)
y^2 e^{2y} \, [I_0(y)K_0(y) - I_1(y)K_1(y)],
\label{disk_vc1}
\end{equation}
where $K = 1.7051\pi$ for a disk with a vertical distribution intermediate between exponential and isothermal.  Since the rotational velocities of our disks are well measured \citep{cth00, fb+07, p+01, dB+08, ZEH90}, and the behavior of $h_R/h_z$ with Hubble type and rotation speed is known from the literature \citep[\eg][]{dG98, kvd02, yd06}, our measurements of $\sigma_z(0)$ place an immediate constraint on the importance of dark matter near the centers of galaxies.  

The result of this analysis is illustrated in Figure~\ref{rotcurves} and summarized in column 7 of Table~\ref{tabhz}.  In brief, the five disks examined in this survey display a range of properties, depending on the Hubble type of the system.  The disks of our latest galaxies, IC~342 and M101, are clearly sub-maximal, requiring scale-length to scale-height ratios $h_R/h_z > 40$ to attain the maximal condition.  Unless these disks are extremely thin (100~pc for IC~342 and 60~pc for M101), their dark matter halos must be contributing significantly to the potential of their inner regions.   In the Sc galaxies M74 and M83, the situation is not as clear.  For M74's disk to be maximal, $h_R/h_z \sim 18$ (\ie\ $h_z \sim 170$~pc).  Although this seems significantly larger than the value of $\sim$8 expected for an Sc galaxy \citep{dG98}, the two values can be forced into marginal agreement if one assumes that the vertical structure of the disk is exponential, rather than intermediate or isothermal.   Similarly, M83 has a requisite scale ratio near the edge of the range expected for SBc galaxies, $h_R/h_z \sim 10$.  Only the Sab galaxy M94, with its high surface-brightness bulge has a disk that can be considered maximal:  if we model the galaxy's inner regions with an $h_z \sim 300$~pc disk and a $\Upsilon_R \sim 1.3$ \citet{h90} bulge ($a = 0.586$~kpc), then the rotation curve of the central $\sim$1.5~kpc is entirely explained by baryons.  We note that with this simple model, the mass-to-light ratio of M94's inner disk is actually greater than that of its bulge; however, if the disk's vertical structure is modeled as isothermal, rather than intermediate, or if the disk's scale height is smaller than $\sim$300~pc, then the mass-to-light ratios come into much better agreement.

Since four of our galactic disks are sub-maximal, it is a simple matter to remove the contribution of their rotation curves and examine the potential of the remaining dark matter component.  When describing halo distributions, two models are generally considered:  one following the predictions of \citet[][hereafter NFW]{NFW96, NFW97} and one obeying a pseudo-isothermal (ISO) law \citep{bs80, kent86}.  The rotation curve for the former is given by
\begin{equation}
V_{NFW}(R) = V_{200}\sqrt{\frac{\ln (1+cx)-cx/(1+cx)}{x[\ln (1+c) - c/(1+c)]}},
\label{NFW}
\end{equation}
where $x = R/R_{200}$, $R_{200}$ is the radius at which the density contrast over the background exceeds 200, $V_{200}$ is the circular velocity at $R_{200}$, and $c$ is a concentration parameter, which physically should always be greater than 1.  In contrast, the rotation curve for the pseudo-isothermal law is 
\begin{equation}
V_{ISO}(R) = \sqrt{4\pi G \rho_0 R_C^2 \left[1-\frac{R_C}{R}
\arctan\left(\frac{R}{R_C}\right)\right]},
\label{ISO}
\end{equation}
where $\rho_0$ is the central density of the halo and $R_C$ is the core radius.  Figure~\ref{rotcurves} fits both these profiles to our residual rotation curves.  As the figure illustrates, the isothermal model fits the data extremely well with mean variances of 0.24, 0.17, 0.04, and 4.52~km$^2$~s$^{-2}$ for IC~342, M74, M101 and M83, respectively.  (In M83, the lack of PN identifications in the central $\sim$1~disk scale length limits the accuracy of our extrapolation.)  In contrast, the NFW fits are all very poor, with large mean variances (39.2, 37.2, 8.51, and 174.8~km$^2$~s$^{-2}$ for these same four galaxies), unphysical concentration indices ($c \ll 1$) and density distributions that overpredict the inner rotation curve.  As in the case of low-surface brightness galaxies \citep[\eg][]{kuzio} and dwarf galaxies \citep[\eg][]{oh+08}, the inner rotation curves of normal spirals are not well represented by NFW profiles.

\subsection{Outer Disks}

In contrast to the galaxies' inner regions, the outer disks of spirals are not constant $\Upsilon$, constant $h_z$ systems.  As the data of M83, M94, and perhaps M74 show, in these distant regions, it is the product of surface mass times scale height that appears constant.  Specifically, in M94, the ratio of $\sigma_z^2 / (G \mu) = K \Upsilon h_z$ has one value in the galaxy's inner regions, and another larger value in the outer disk, and in M83, the ratio blows up at large radii.  (The same effect can be seen in the outer regions of M33 \citep{M33PNe}, though the larger inclination of this galaxy creates an ambiguity in the interpretation.)  At radii greater than $\sim$3~disk scale lengths, the disks of spiral galaxies are either flaring, increasing their mass-to-light ratio, or changing their luminosity profile from that of a single scale-length exponential.

To some extent, all three of these phenomena must be happening.  Since many galaxies have large H{\sc i} envelopes, an increase in the disk mass-to-optical light ratio at large radii is not unexpected.  However, measurements from the THINGS survey \citep{THINGS} demonstrate that this effect is minor:  even in the outer regions of M83 and M94, H{\sc i} contributes no more than $\sim$10\% of the kinematically-estimated disk mass.   Similarly, many spirals do not have simple exponential disks in their outer regions.  In M94, a shallowing of the surface brightness profile was recorded by \citet{g+09} and SINGS \citep{dB+08}, and this excess luminosity goes a long way towards explaining the system's mass-to-light ratio.  Unfortunately, not all spirals have such breaks, and the disk of M83 appears to follow the same exponential over the entire range of our survey.   Finally, stability arguments demand that some flaring occurs in the outer parts of spirals.   In all of our galaxies (except IC~342, whose data only extends to $\sim$2~disk scale lengths), $h_z$ likely increases past $\sim$3 disk scale lengths, in order to satisfy the \citet{t64} criterion against axisymmetric perturbations.

Is there any direct evidence for flaring in the outer disks of spirals?  Studies of edge-on galaxies have generally found that galactic scale heights are roughly constant with radius \citep[\eg][]{vdKs82, bm02}.  Even when radial gradients have been found, they have generally been small, and linked to the presence of the thick disks of early-type systems \citep{dGP97}.  However, these studies have mostly been confined to the systems' bright inner regions.  In NGC~891, which has extremely deep surface photometry derived from individual star counts, a thick, highly-flattened, low surface brightness structure does extend well beyond the limits of the galaxy's thin disk \citep{mouhcine}.

Mechanisms do exist which can cause the scale height of stars to increase rapidly at large radii.  \citet{nj02} have shown that self-gravitating disks acting under the influence of H{\sc i} and H$_2$ gas and a dark matter halo can develop moderately large scale height gradients in their outer regions ($R \gtrsim 2$~disk scale lengths).  Alternatively, flared-disks and higher than expected values of $\sigma_z$ can be caused by interactions with halo substructures.  In fact, Model~F by \citet{hc06}, which simulates the heating of the Milky Way's disk by a population of $\sim$300 moderately massive ($\sim$10$^{8.5} \, M_{\odot}$) dark matter subhalos, produces a velocity dispersion profile that is an excellent match to that seen in M83 and M94 \citep{Letter}.  Recent simulations by S. Kazantzidis (2009, private communications) confirm this result, as they reproduce the full $\sigma_z$ profiles of the two galaxies extremely well.  Thus, the flaring inferred by our kinematic data is certainly plausible.

\section{CONCLUSIONS}

An analysis of 550 precise PN velocities has yielded kinematic mass estimates throughout the disks of five low-inclination spiral galaxies.  In the inner regions of IC~342, M74, and M94, we find that the velocity dispersion perpendicular to the galactic disk exponentially decays with a scale length twice that of the optical light.  While our observations of the remaining two galaxies yield ambiguous results (due to a Lindbland resonance in M83 and a poorly sampled velocity field in M101), the data are consistent with expectations from a constant scale height, constant mass-to-light ratio disk.   In the later-type (Scd) systems, these disks are clearly sub-maximal, with surface mass densities less than a quarter that needed to reproduce the central rotation curves.  In earlier (Sc) galaxies, this discrepancy is smaller, but still present; only the very-early Sab system M94 has evidence for a maximal disk.  

When we subtract off the disk contribution from the H{\sc i} rotation curve, we infer the potential of the galaxies' invisible dark halos in all our galaxies except M94.  (The uncertainties in the contribution of M94's bright bulge preclude such an analysis in this earliest galaxy in our sample.)   In all four cases, these halos are best fit with a simple pseudo-isothermal model; NFW profiles are poor fits to the data, as they always overpredict the residual rotation curve.  As in the analysis of low surface-brightness and dwarf galaxies, we find no evidence for central dark matter cusps in our systems.

Interestingly, once outside of $\sim$3~disk scale lengths, the $z$-direction velocity dispersion of galactic disks no longer declines, but instead stays constant with radius.  This behavior, coupled with stability arguments, suggests that the disks of spiral galaxies flare substantially at extremely large radii, and may have strongly increasing mass-to-light ratios.   Whether this flaring is related to the increased importance of the thick disk, or heating of the thin disk by halo substructure, is unclear at the present time.  To further investigate this phenomenon, deep surface photometry of the extreme outer disks of edge-on galaxies is needed.

Our kinematic mass estimates for galactic disks are consistent with expectations.  The galaxies with the latest Hubble types (IC~342, M101) have the smallest mass-to-light ratios, while earlier systems (M74 and M94) have values of $\Upsilon$ greater than one.  Unfortunately, at this time, no deep, multi-band optical and IR surface photometry exists for any of our galaxies.  Once these data are acquired, we will be able to compare our kinematic values for $\Upsilon$ to stellar mass-to-light ratio estimates from population synthesis models.  Such a confrontation will not only produce a better constraint on galactic populations, but also will improve our knowledge of the phase-space distribution of galactic disks.

\acknowledgments
We thank Steinn Sigurdsson for very useful discussions, and Fabian Walter, Erwin de Blok, and the entire THINGS collaboration for sharing their H{\sc i} maps prior to publication.  We also acknowledge the helpful suggestions of our anonymous referee.  This research has made use of NASA's Astrophysics Data System and the NASA/IPAC Extragalactic Database (NED) which is operated by the Jet Propulsion Laboratory, California Institute of Technology, under contract with the National Aeronautics and Space Administration.  This work was supported by NSF grant AST 06-07416 and a Pennsylvania Space Grant Fellowship.

\end{document}